\documentclass[twocolumn]{aastex631}

\usepackage{graphicx}	
\usepackage{amsmath}	
\usepackage{amssymb}
\usepackage{bm}	
\usepackage{lipsum}
\usepackage{afterpage}
\usepackage{multirow}
\usepackage{makecell}

\newcommand{\src}{ESO~511$-$G030}
\newcommand{\nustar}{\textit{NuSTAR}}
\newcommand{\xmm}{\textit{XMM-Newton}}

\begin{document}

\title{The Low Temperature Corona in \src\ Revealed by \nustar\ and \xmm}

\correspondingauthor{Cosimo Bambi} 
\email{bambi@fudan.edu.cn}

\author{Zuobin Zhang}
\affiliation{Center for Field Theory and Particle Physics and Department of Physics, Fudan University, 2005 Songhu Road, 200438 Shanghai, China}

\author[0000-0002-9639-4352]{Jiachen Jiang}
\affiliation{Institute of Astronomy, University of Cambridge, Madingley Road, Cambridge CB3 0HA, UK}

\author{Honghui Liu}
\affiliation{Center for Field Theory and Particle Physics and Department of Physics, Fudan University, 2005 Songhu Road, 200438 Shanghai, China}

\author[0000-0002-3180-9502]{Cosimo Bambi}
\affiliation{Center for Field Theory and Particle Physics and Department of Physics, Fudan University, 2005 Songhu Road, 200438 Shanghai, China}

\author{Christopher S. Reynolds}
\affiliation{Institute of Astronomy, University of Cambridge, Madingley Road, Cambridge CB3 0HA, UK}

\author{Andrew C. Fabian}
\affiliation{Institute of Astronomy, University of Cambridge, Madingley Road, Cambridge CB3 0HA, UK}

\author{Thomas Dauser}
\affiliation{Remeis-Observatory \& ECAP, FAU Erlangen-N\"urnberg, Sternwartstr. 7, 96049 Bamberg, Germany}

\author{Kristin Madsen}
\affiliation{Cahill Center for Astronomy and Astrophysics, California Institute of Technology, Pasadena, CA 91125, USA}
\affiliation{CRESST and X-ray Astrophysics Laboratory, NASA Goddard Space Flight Center, Greenbelt, MD 20771 USA}

\author{Andrew Young}
\affiliation{School of Physics, Tyndall Avenue, University of Bristol, Bristol BS8 1TH, UK}

\author{Luigi Gallo}
\affiliation{Department of Astronomy and Physics, Saint Mary’s University, 923 Robie Street, Halifax, NS B3H 3C 3, Canada}

\author{Zhibo Yu}
\affiliation{Center for Field Theory and Particle Physics and Department of Physics, Fudan University, 2005 Songhu Road, 200438 Shanghai, China}

\author{John Tomsick}
\affiliation{Space Sciences Laboratory, University of California, 7 Gauss Way, Berkeley, CA 94720-7450, USA}

\begin{abstract}

We present the results from a coordinated \xmm\ $+$ \nustar\ observation of the Seyfert 1 Galaxy \src. With this joint monitoring programme, we conduct a detailed variability and spectral analysis. The source remained in a low flux and very stable state throughout the observation period, although there are slight fluctuations of flux over long timescales. The broadband (0.3–78~keV) spectrum shows the presence of a power-law continuum with a soft excess below 2~keV, a relatively narrow iron K$\alpha$ emission ($\sim$6.4~keV), and an obvious cutoff at high energies. We find that the soft excess can be modeled by two different possible scenarios: a warm ($kT_{\rm e} \sim$ 0.19~keV) and optically thick ($\tau - 18\sim25$) Comptonizing corona or a relativistic reflection from a high-density ($\log [n_{\rm e}/{\rm cm}^{-3}]=17.1 \sim 18.5$) inner disc. All models require a low temperature ($kT_{\rm e} \sim$ 13~keV) for the hot corona.

\end{abstract}


\section{Introduction} \label{sec:intro}

Active galactic nuclei (AGNs) are luminous sources in the Universe over a broad energy range from radio to gamma rays. This emission results from the accretion of matter onto a supermassive black hole (SMBH) at the center of its host galaxy \citep{Lynden-Bell1969,Rees1984}. By studying the X-ray emission, we can directly probe the innermost region of AGNs. The AGN X-ray emission region is small in size \citep{Reis2013} and located close to the central SMBH and accretion disc, as suggested by studies of black hole mass dependence of AGN X-ray variability (e.g., \citealt{Axelsson2013}; \citealt{McHardy2013}; \citealt{Ludlam2015}), reverberation of X-ray radiation reprocessed by the accretion disc (e.g., \citealt{DeMarco2013}; \citealt{Uttley2014}; \citealt{Kara2016}), and quasar microlensing (e.g., \citealt{Mosquera2013}; \citealt{Chartas2016}; \citealt{Guerras2017}).

The typical broadband X-ray spectrum of a Seyfert~1 AGN consists of a power-law continuum, fluorescent emission lines, a Compton hump, and a soft excess below $\sim$2 keV. In the simplest scenario, thermal emission from the disc mostly emits in the ultraviolet (UV) band. The seed UV/optical photons are Compton up-scattered to the hard X-ray band in a region filled with a hot and optically thin plasma near the central black hole, which is often referred to as the corona (e.g., \citealt{Vaiana1978};  \citealt{Haardt1991}; \citealt{Merloni2003}). A fraction of the hard X-ray photons illuminate the surface of the accretion disc and are reflected to produce the reflection component (e.g., \citealt{Ross2005}; \citealt{Garcia2010}, which is smeared by relativistic effects (e.g., \citealt{2021SSRv..217...65B, Fabian1989}). 

Broadband X-ray spectroscopy of the X-ray emission produced in the Comptonizing plasma can provide important insights into the principal properties of the corona, such as its temperature ($kT_{\rm e}$), optical depth ($\tau_{\rm e}$), and ultimately its geometry. The continuum originates in the Comptonization processes of low-energy disc photons scattered by hot electrons (e.g., \citealt{Balokovic2020}). The high-energy turnover is generally interpreted as the temperature of the corona (or a value close to it). Recently, \cite{Fabian2015} gathered results from \textit{NuSTAR} to map out their locations on the compactness-temperature ($\ell - \Theta$) diagram, and found many sources are (marginally) above the electron–electron coupling line and all are above the electron–proton line. Then \cite{Fabian2017} re-examined the case of hybrid coronae \citep{Zdziarski1993}, where the plasma contains both thermal and non-thermal particles, and found that objects with the lowest coronal temperature measurements require the largest non-thermal fractions.

\src\ is a bare Seyfert 1 AGN  at redshift $z = 0.0224$ (\citealt{Tombesi2010}, \citealt{Winter2012}, and \citealt{Laha2014}). It is found to be one of the brightest bare Seyfert AGNs featured in the Swift 58 month BAT catalog \citep{Winter2012}. \cite{Ghosh2021} presented a broadband (optical-UV to hard X-ray) spectral study of \src\, using multi-epoch \textit{Suzaku} and \xmm\ \citep{Jansen2001} data from 2007 and 2012. They investigated the spectral features observed in the source with a physically motivated set of models and reported a rapidly spinning black hole ($a_* > 0.78$) and a compact corona, indicating a relativistic origin of the broad Fe emission line. \cite{Ghosh2021} also found an inner disc temperature of $2\sim3$~eV, which characterizes the UV bump, and that the SMBH  accretes at a sub-Eddington rate ($\lambda_{\rm Edd}=0.004-0.008$).

Previous studies of \src\ were based on the \xmm\ observation data in 2007 and the \textit{Suzaku} observation in 2012 (e.g., \citealt{Ghosh2021}, \citealt{Tombesi2010}, \citealt{Winter2012}, and \citealt{Laha2014}). \nustar\ \citep{Harrison2013} and \xmm\ conducted a joint observing campaign on this source in 2019, consisting of five joint observations over twenty days. In this paper, we analyze the observational data from this \nustar\ and \xmm\ joint observation campaign, investigating the nature of the corona continuum, reflection, and soft excess.

The paper is organized as follows. In Sec.~\ref{observations}, we present the observational data reduction and the light curves. The spectral analysis with two different possible scenarios is reported in Sec.~\ref{analysis}. We discuss the results and report our conclusions in Sec.~\ref{discussion} and Sec.~\ref{conclusion}, respectively.


\section{Observations and data reduction}
\label{observations}

During the period of 07-20-2019 to 08-09-2019, \nustar\ and \xmm\ performed five joint observations. The details of these observations are in Tab.~\ref{t-obs}. These data are taken with two focal plane modules, named FPMA and FPMB, on board the \nustar\ satellite, and EPIC-pn CCD modules, on board of \xmm. A total unfiltered exposure time of $\sim 168$ ks with five different observations is obtained.


\begin{table*}
\centering
\caption{\rm Summary of the observations analyzed in the present work. \label{t-obs}}
\renewcommand\arraystretch{1.5}
\begin{tabular}{m{1.5cm}m{2.5cm}<{\centering}m{2.5cm}<{\centering}m{2.5cm}<{\centering}m{2.5cm}<{\centering}m{2.5cm}<{\centering}}
\hline\hline

&  Mission & Obs.~ ID & Instrument &  Start data &  Exposure (ks)  \\ \hline
Epoch~1 & \nustar\       & 60502035002 & FPMA/B       & 2019-07-20            &      32.1      \\   
           &\xmm\        & 0852010101  & EPIC-pn      & 2019-07-20 &      37.0 \\ \hline
Epoch~2 & \nustar\       & 60502035004 & FPMA/B       & 2019-07-25            &      34.1        \\ 
           &\xmm\        & 0852010201  & EPIC-pn      & 2019-07-25 &      36.0  \\ \hline
Epoch~3 & \nustar\       & 60502035006 & FPMA/B       & 2019-07-29            &      31.2       \\ 
           &\xmm\        & 0852010301  & EPIC-pn      & 2019-07-29 &      33.0       \\ \hline
Epoch~4 & \nustar\       & 60502035008 & FPMA/B       & 2019-08-02            &      41.8        \\
           &\xmm\        & 0852010401  & EPIC-pn      & 2019-08-02 &      38.3        \\ \hline
Epoch~5 & \nustar\       & 60502035010 & FPMA/B       & 2019-08-09            &      29.2       \\ 
          &\xmm\         & 0852010501  & EPIC-pn      & 2019-08-09 &      36.2       \\ 
\hline\hline
\end{tabular}
\vspace{0.3cm}
\end{table*}


\subsection{\xmm\ data reduction}

The \xmm\ EPIC cameras offer the possibility of performing extremely sensitive imaging observations over a field of view of $30'$ and the energy range from 0.2 to 12 keV, with moderate spectral ( $E/\Delta E \sim 20-50$) and angular resolution ($\sim 6''$ FWHM; $\sim 15''$ HEW). Because of the higher effective area of EPIC-pn compared with EPIC-MOS and its consistency with EPIC-MOS data, we only consider EPIC-pn data in the 0.3–10.0 keV energy band in our X-ray spectral analysis.

We reduce the data and extract products from Observation Data Files (ODF) following the standard procedures based on the XMM-Newton Science Analysis System (SAS 18.0.0) and the latest calibration files. The EPIC-pn data are produced using \texttt{epproc} and processed with the standard filtering criterion. Then we remove periods of high background by creating a Good Time Interval (GTI) file using the task \texttt{tabgitgen}. The source products are extracted from a circular region with a radius of $30''$ centered on the source, and the background is taken from a circular region with a radius of $65''$ offset source. The \texttt{evselect} task was used to select single and double events for EPIC-pn (PATTERN $\leq$ 4, FLAG $==$ 0) source event lists. The Redistribution Matrix File (RMF) and Ancillary Response File (ARF) are created by using the SAS tasks \texttt{rmfgen} and \texttt{arfgen}, respectively. We test the pile-up for these data by using the SAS task \texttt{epatplot} and find that the influence of pile-up events is negligible during the observations.  

\subsection{\nustar\ data reduction}

The reduction of the \nustar\ \citep{Harrison2013} data was conducted following the standard procedure using the NuSTAR Data Analysis Software (NUSTARDAS v.2.1.1), and updated calibration files from \nustar\ CALDB v20220301. We produce calibrated and filtered event files with \texttt{nupipeline}. Passages through the South Atlantic Anomaly are excluded from consideration using the following settings: saamode = STRICT,  SAACALC = 2 and TENTACLE = YES. For such an AGN source, the effect of the SAA filtering is non-negligible. To get more reliable spectra products, we use strict filtering criteria here to realize the reduction in the background rates.  

\begin{figure}
    \centering
    \includegraphics[width=0.80\linewidth, trim=140 20 110 20, clip]{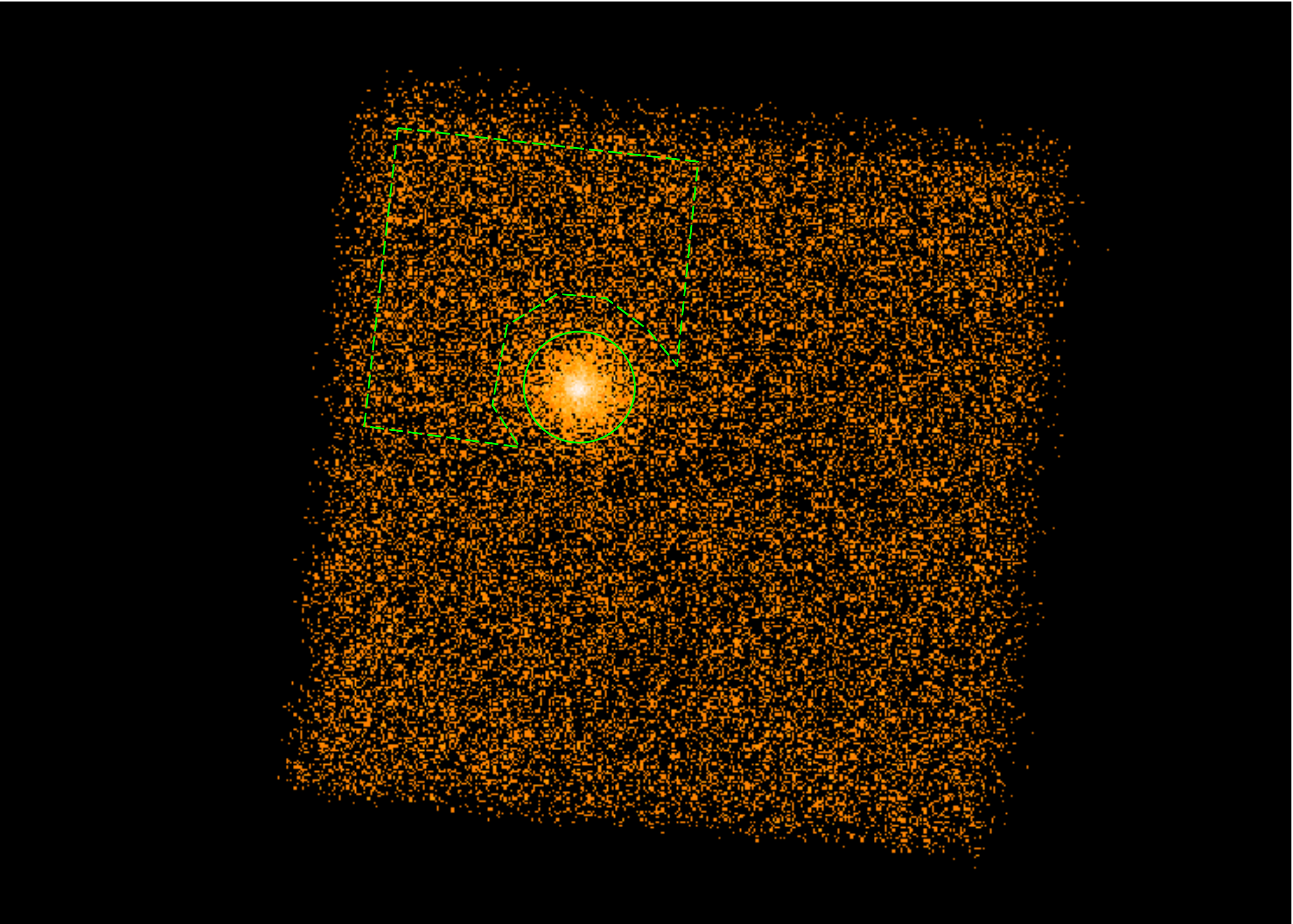}
    \caption{\nustar\ FPMB image of observation ID 60502035008 (\src). The source spectrum is extracted from the solid green circle region with a radius of $65''$ and the background spectrum is extracted from the dashed green polygon.}
    \label{image}
\end{figure}

We utilize the task package \texttt{nuproduct} to extract source products and their associated instrumental response files from a circular region of radius $65''$ centered on the source. The \nustar\ background varies across the field of view and between the four CdZnTe (CZT) detectors on each focal plane. For the background, we extract it from an optional maximal nearby polygon region free from source contamination, and we limit the background region on the CZT chip of the source, as shown in Fig.~\ref{image}.  With these strategies, we minimize the systematic errors from the instrument and obtain a more reliable background spectrum. The products are obtained from FPMA and FPMB separately.


\subsection{Lightcurves and variability}

\begin{figure*}
    \centering
    \includegraphics[width=0.32\linewidth]{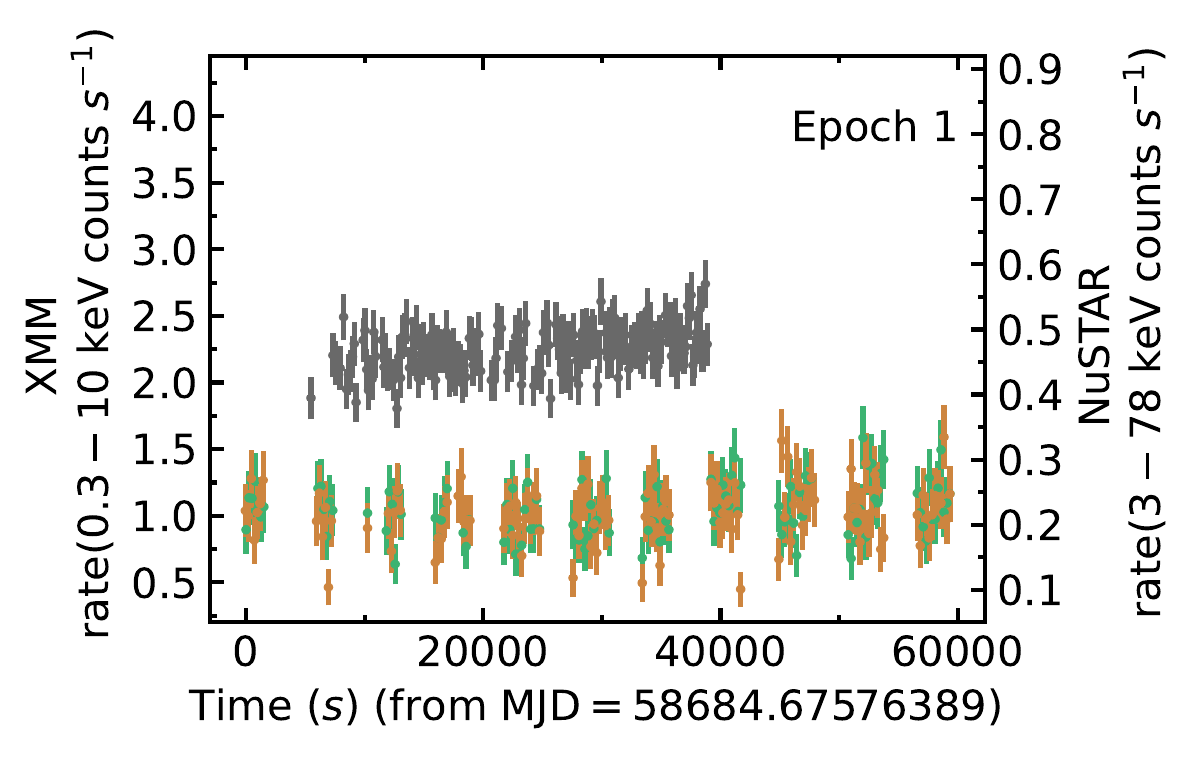}
    \includegraphics[width=0.32\linewidth]{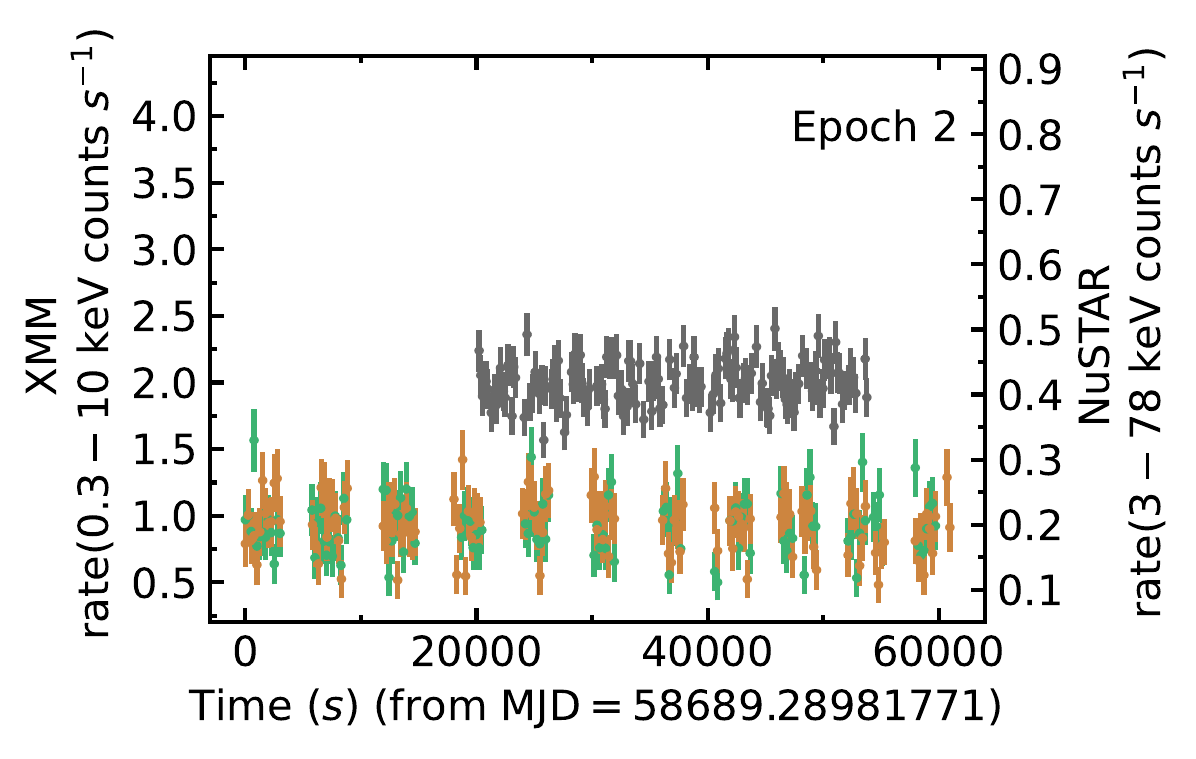}
    \includegraphics[width=0.32\linewidth]{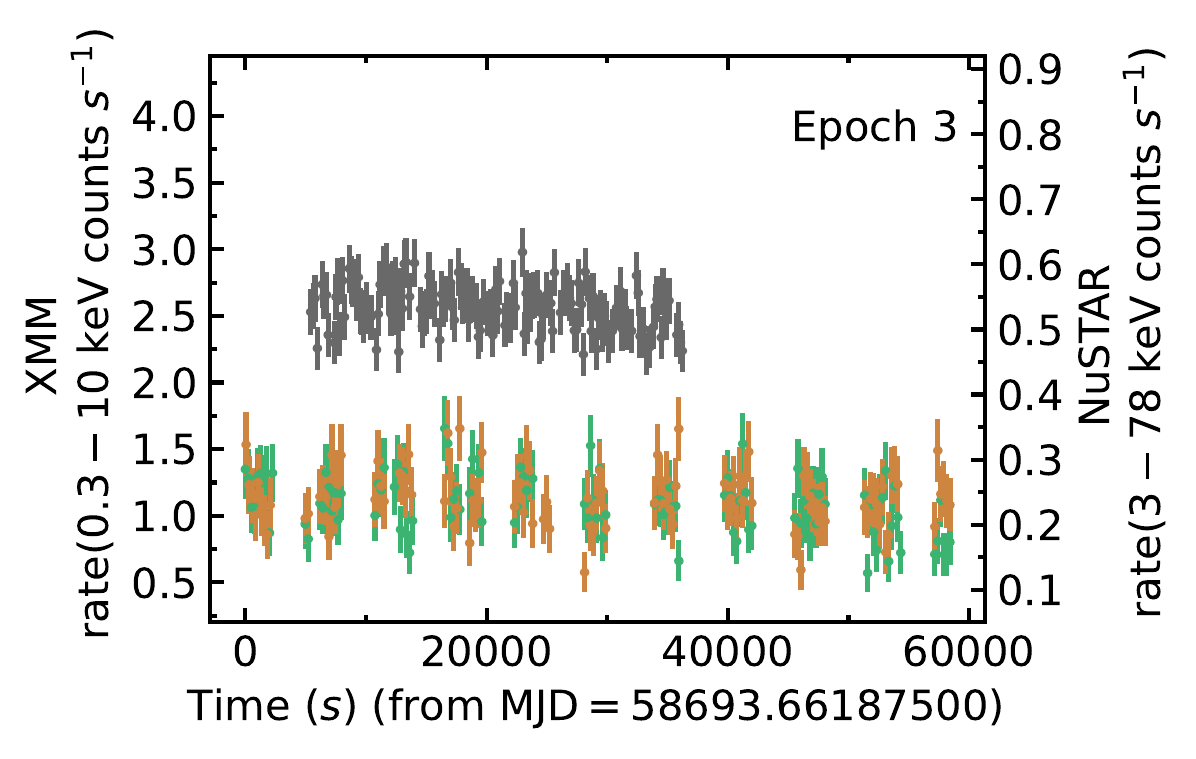} \\
    \includegraphics[width=0.32\linewidth]{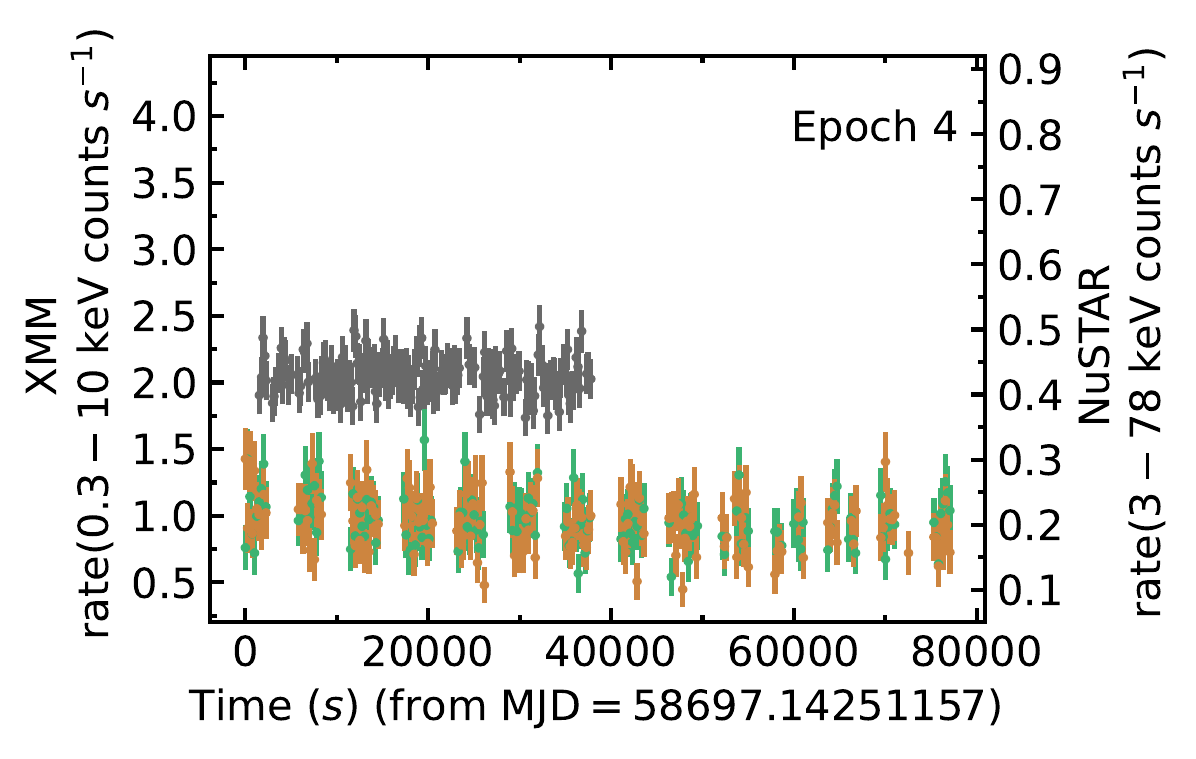}
    \includegraphics[width=0.32\linewidth]{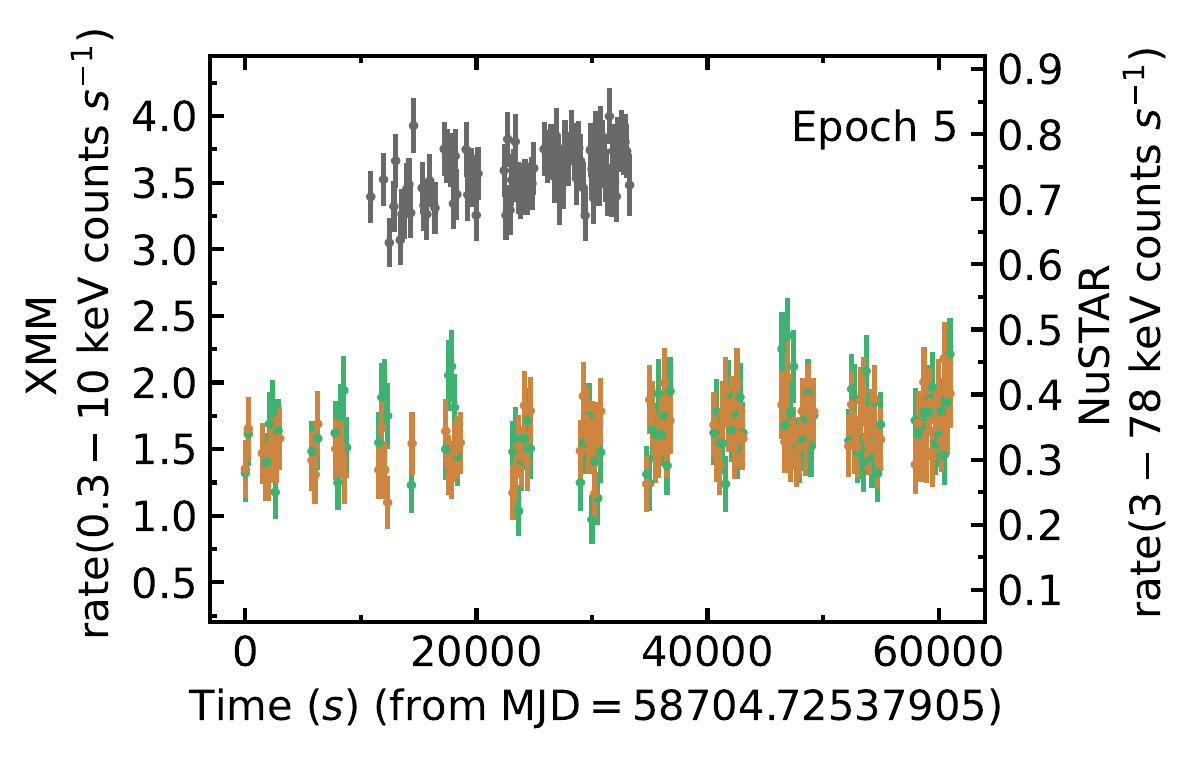} \\
    \caption{Light curves of \nustar\ and simultaneous \xmm\ observations of \src. The \xmm\ data are binned in 150~s intervals and shown in grey. The \nustar\ FPMA and FPMB data are binned in 200~s intervals and shown in orange and green, respectively. \src\ remains in a consistent flux state during each observation.}
    \label{lcurve-1}
\end{figure*}

Fig.~\ref{lcurve-1} presents the light curves of \nustar\ and simultaneous \xmm\ observations on Epoch 1-5. \xmm\ data are binned in 150~s intervals and \nustar\ FPMA data and FPMB data are binned in 200~s intervals.  The \xmm\ $0.3 - 10$~keV count rate of \src\ remains consistent within the range of $1.8 - 3.0$~ct~s$^{-1}$ during the first four epochs, and \nustar\ $3.0 - 78.0$~keV count rate within the range of $0.1 - 0.35$ ct s$^{-1}$. In Epoch 5, the light curve shows an increase of $\sim60$\% both in \nustar\ and \xmm\ count rates, compared to the average count rates of Epoch~1-4. In Epoch 5, which is the brightest epoch, the \xmm\ light curve shows a count rate of $\sim 3.5$ counts/s and \nustar\ shows a count rate of $\sim 0.35$ counts/s. In Epoch 2, which is the faintest epoch, the \xmm\ light curve shows a count rate of $\sim 2.0$ counts/s and \nustar\ shows a count rate of $\sim 0.2$ counts/s. Note that the photon rate fluctuation between different observations occurs simultaneously in multiple instruments, indicating that the variability occurs simultaneously for the entire broad energy band.

\begin{figure*}
    \centering
    \includegraphics[width=0.32\linewidth]{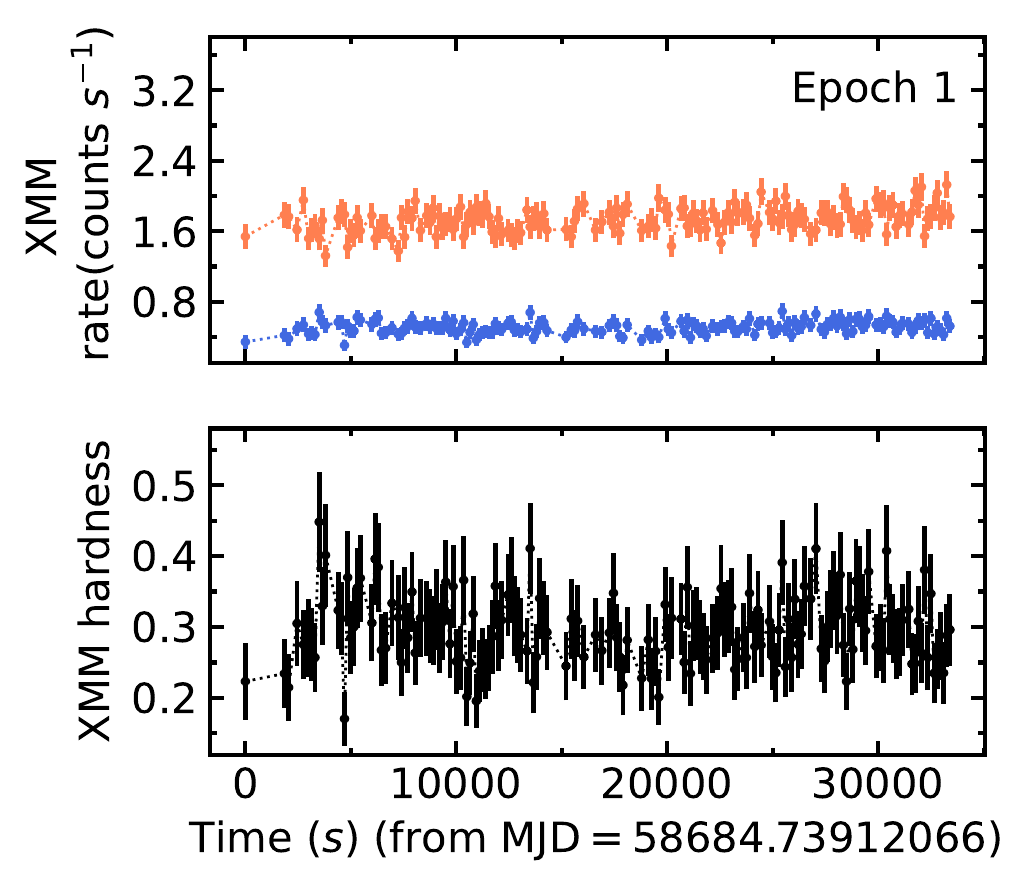}
    \includegraphics[width=0.32\linewidth]{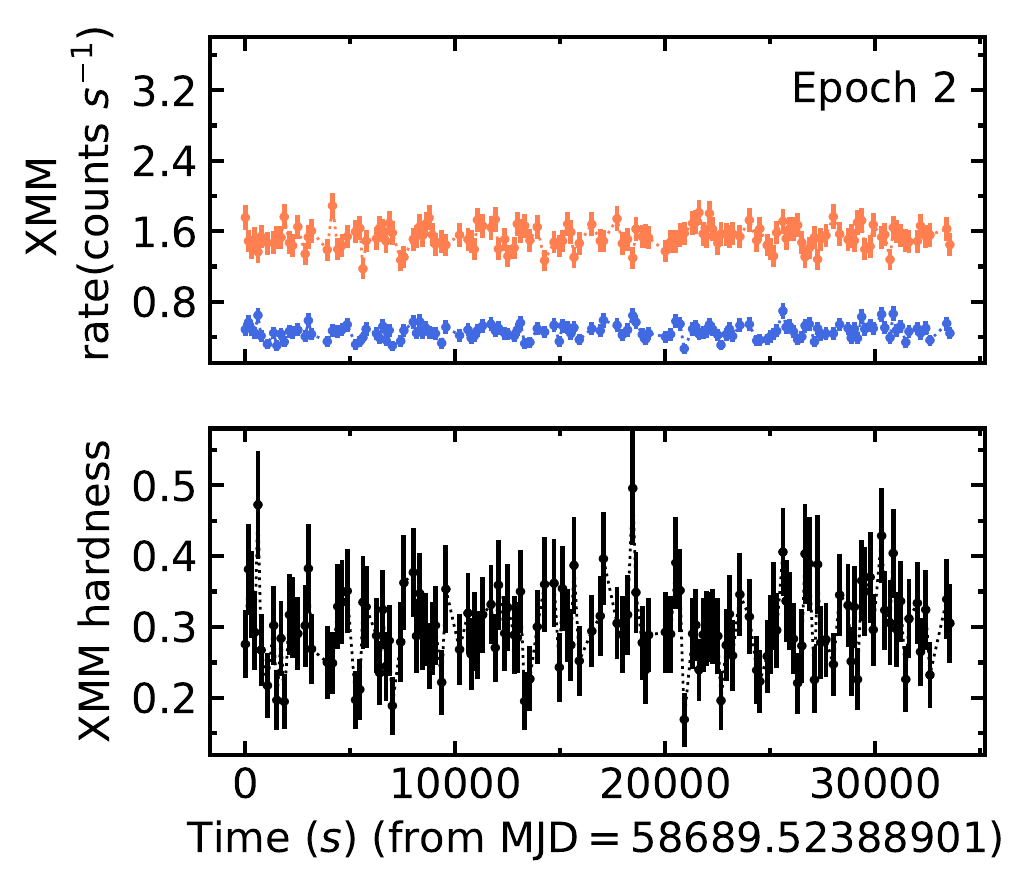}
    \includegraphics[width=0.32\linewidth]{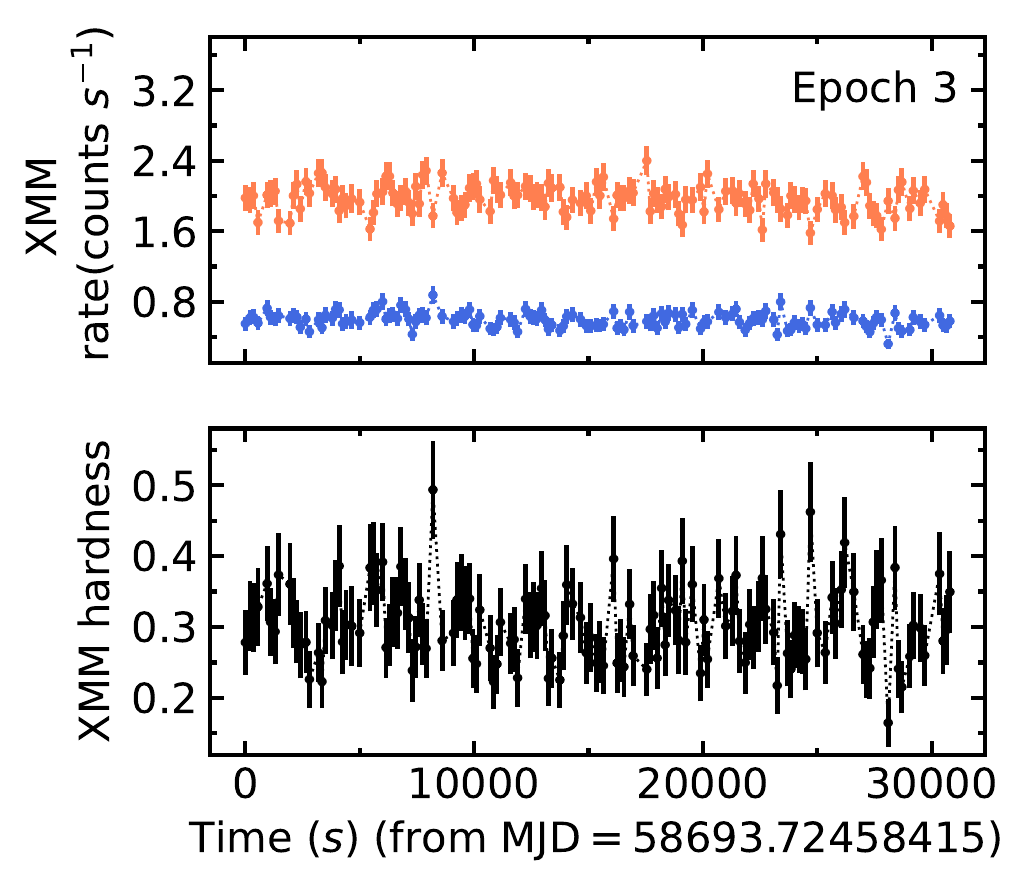} \\
    \includegraphics[width=0.32\linewidth]{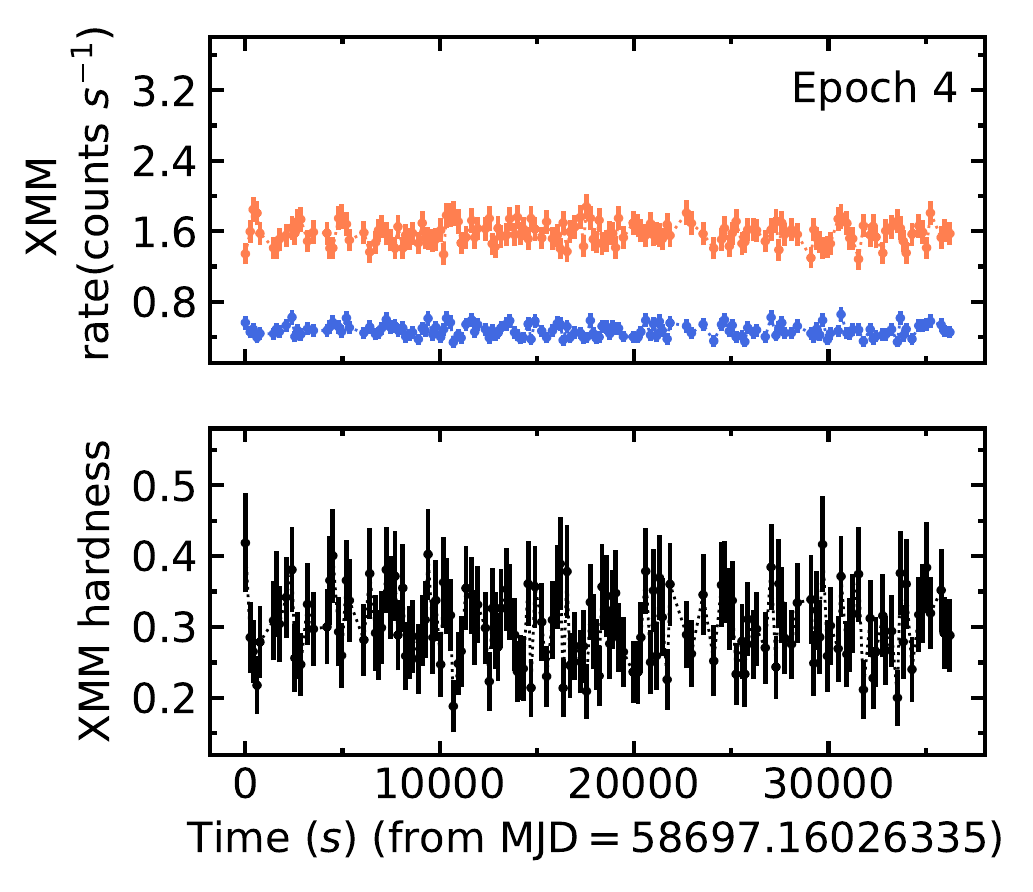}
    \includegraphics[width=0.32\linewidth]{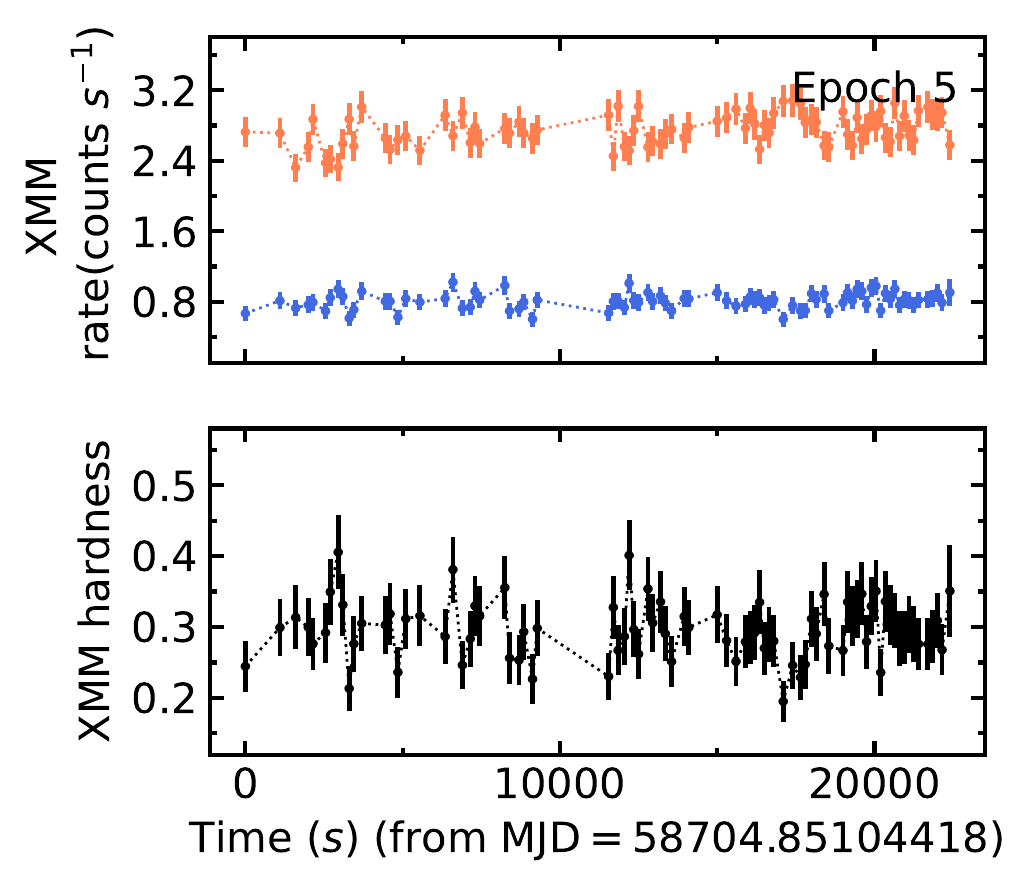} \\
    \caption{The light curves and corresponding hardness ratios extracted from the \xmm\ observations in Epoch 1-5. The upper panels show the EPIC-pn light curves in the 0.3–2.0 keV (red) and 2–10 keV (blue) bands. The lower panels show the corresponding hardness ratios (2.0–10.0/0.3–2.0 keV). The bin time for the light curve is set to 150 s. The hardness ratio remains consistent between 0.2 and 0.4 in all epochs.}
    \label{lcurve-2}
\end{figure*}

To investigate this variability, we extract the \xmm\ light curves in the 0.3–2.0 keV and 2–10 keV bands, shown in the upper panel of each plot in Fig.~\ref{lcurve-2}, which shows that both soft and hard energy bands vary simultaneously. Moreover, we plot the \xmm\ hardness (2–10 keV/0.3–2.0 keV) ratio in the lower panel of each plot. The hardness presents a stable trend both in a single observation and between observations. We also extract the spectra from divided epochs and only a tiny discrepancy in normalization is found between spectra, which confirms the result obtained from the light curve analysis. 

As mentioned above, because variability occurs simultaneously for the entire broad energy band and the spectrum variations are only a discrepancy in normalization between the observations, we merge the spectra of Epoch 1-5 into a single spectrum. For \xmm, we produce a combined multi-observation EPIC-pn spectrum using the SAS ftool \texttt{epicspeccombine} and we focus on the \xmm\ data over the 0.3–10.0 keV band in the following analysis. For \nustar, we produce a combined multi-observation FPMA spectrum and FPMB spectrum using the HEASARC ftool \texttt{addspec} and we use the \nustar\ data over the 3.0–78.0 keV band. All spectra are rebinned to minimum counts of 20 per energy bin and oversample the spectral resolution by a factor of 3.

\section{Spectral Analysis}
\label{analysis}

In this section, we present an analysis of the time-averaged \nustar\ and \xmm\ spectra using the XSPEC (v12.12.1) package \citep{Arnaud1996}. To account for the differences between the detector responses of FPMA/B and EPIC-pn, we include a cross-calibration factor, which is fixed to unity for the EPIC-pn spectra, but varies freely for FPMA and FPMB \citep{Madsen2015a}. The $\chi^{2}$ statistics is employed and all parameter uncertainties are estimated at 90\% confidence level, corresponding to $\Delta \chi^{2}=2.71$. We include the absorption model \texttt{tbnew} to describe the Galactic absorption, using the recommended photoelectric cross sections of \citet{Verner1996}. In addition, the multiplicative model \texttt{zmshift} is used to account for the redshift of the source and fix the redshift at $z=0.0224$ during the spectral fitting.

\begin{figure}
    \centering
    \includegraphics[width=0.9\linewidth]{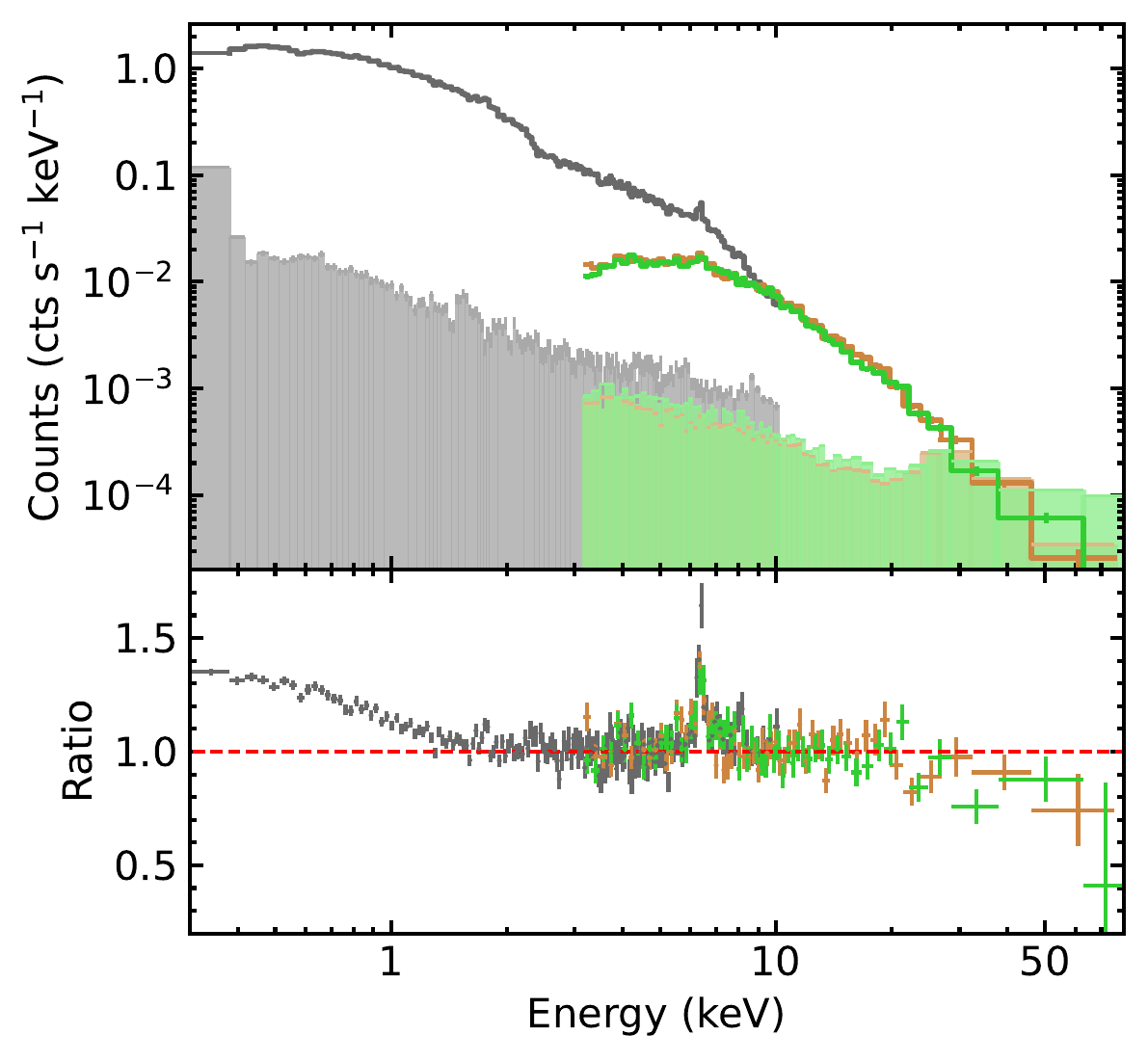} \\
    \caption{{Top: \nustar\ (orange: FPMA; green: FPMB) and \xmm\ (gray: EPIC-pn) source (crosses) and background spectra (shaded regions) of \src .} Bottom: data-to-model ratio using an absorbed power-law model: grey for \xmm\ data, orange and green for \nustar\ FPMA and FPMB data, respectively. The spectrum is fitted ignoring the data below 3 keV, above 15 keV and the 5-7 keV band (i.e., ignoring the soft excess region, the iron emission band, and Compton hump). The spectrum of \src\ shows evidence of strong soft excess emission below 2 keV,  iron K emission line at $\sim$6.4 keV, and a cutoff at high energies. The data have been rebinned for visual clarity.}
    \label{soft_excess_iron}
\end{figure}

We start the fitting with an absorbed power-law model, i.e. \texttt{tbnew $\times$ powerlaw} in XSPEC language. \texttt{powerlaw} accounts for a power-law component from the corona. In this fit, we ignore the data below 3 keV, above 15 keV and the 5-7 keV band, i.e., we ignore the possible soft excess, the iron emission line, and Compton hump. Fig.~\ref{soft_excess_iron} shows the broadband spectra of \src\ (upper panel) and the extrapolated data-to-model ratios of the 0.3-78.0 keV dataset from the above fit. The typical AGN spectral features can be seen: a soft excess below 2 keV, Fe K$\alpha$ emission at $\sim$6.4 keV, a weak Compton hump peaking at $\sim$20 keV, and a cutoff at high energy. 

\begin{figure}
    \centering
    \includegraphics[width=0.9\linewidth]{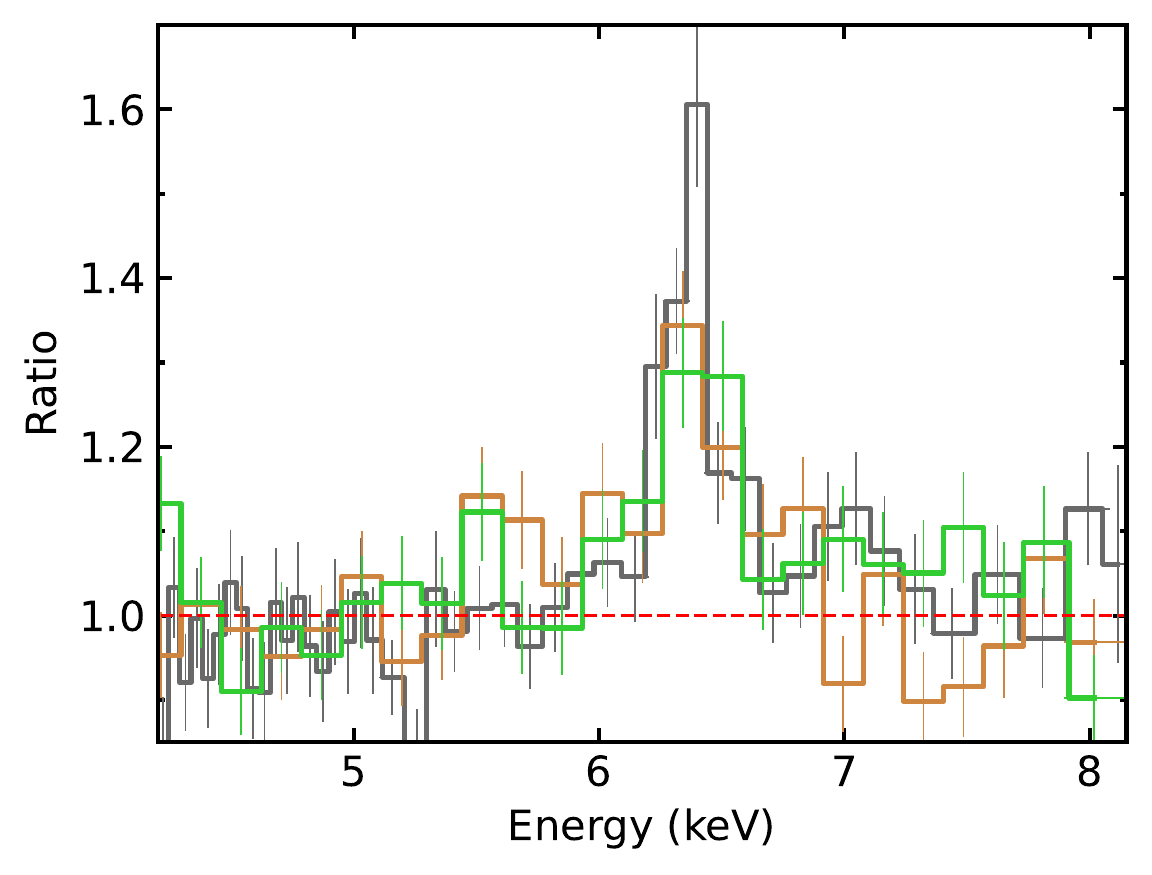} \\
    \caption{A detailed view of the region near the iron K emission line at $6.4$~keV in Fig.~\ref{soft_excess_iron}. As in Fig.~\ref{soft_excess_iron}, the \xmm\ data are shown in grey, \nustar\ FPMA data in orange and \nustar\ FPMB data in green. The \xmm\ and \nustar\ spectra show a consistent shape for the narrow iron K-shell emission line. The data have been rebinned for visual clarity.}
    \label{iron_line}
\end{figure}

Fig.~\ref{iron_line} presents a zoomed-in version of the residual in the 6 keV region. Both the \xmm\ and \nustar\ spectra reveal a consistent  shape for the iron K-shell emission line. To investigate the excess emission at 6$\sim$7~keV, we first introduce a \texttt{gaussian} model to the absorbed power-law model. The source frame line energy is consistent with 6.4 keV and line width $\sigma=0.04_{-0.03}^{+0.05}$~keV, which indicates the presence of a relatively narrow Fe emission line in \src. These features can be partially accounted for by a reprocessing of X-ray photons in a neutral and distant material, free from relativistic effects, possibly in the broad-line region (e.g., \citealt{Costantini2016}; \citealt{Nardini2016}), or the torus (e.g., \citealt{Yaqoob2007}; \citealt{Marinucci2018}). To probe the cutoff at high energies, we replace \texttt{powerlaw} with \texttt{nthcomp} (\citealt{Zdziarski1996}; \citealt{Zycki1999}). This model reveals a low temperature corona, $kT_{\rm e}=16_{-4}^{+5}$~keV. In our subsequent analysis, we will investigate it with more physical models.

To fit the soft excess, we separately try a phenomenological single temperature blackbody model, a warm corona model, and a high-density relativistic reflection model. The fits with these models are presented and discussed in the following subsections. The models are summarized in Tab.~\ref{t-mod}. The data-to-model ratios of the fits are depicted in the left column of Fig.~\ref{ratio_chi} and the best-fit models are shown in Fig.~\ref{eemod}. The best-fit results are summarized in Tab.~\ref{best-fit-1}, where $\nu$ is the number of degrees of freedom (dof) and $\chi_{\rm red}^{2}=\chi^{2}/\nu$ is the reduced $\chi^{2}$. Instead of the normalization of every component, in Tab.~\ref{best-fit-1} we report the flux of every component calculated by \texttt{cflux} over the energy range 0.3-80.0 keV.

\begin{table}
\centering
\caption{\rm Summary of the models used in our analysis: blackbody, warm corona, and relativistic reflection. 
\label{t-mod}}
\renewcommand\arraystretch{1.5}
{\scriptsize
\begin{tabular}{lc}
\hline\hline
\makebox[0.024\textwidth]{Model} & Component \\
\hline
1 & \texttt{tbnew}$\times$\texttt{zmshift}$\times$(\texttt{bbody}+\texttt{nthcomp}+\texttt{xillverCp}) \\
2 & \texttt{tbnew}$\times$\texttt{zmshift}$\times$(\texttt{nthcomp}+\texttt{nthcomp}+\texttt{xillverCp}) \\
3 & \texttt{tbnew}$\times$\texttt{zmshift}$\times$(\texttt{relconv}$\times$\texttt{xillverDCp}+\texttt{nthcomp}+\texttt{xillverCp}) \\
\hline\hline
\end{tabular}
}
\vspace{0.5cm}
\end{table}


\subsection{Model 1: phenomenological blackbody model}

In this fit, we use the single temperature blackbody model \texttt{bbody} to describe the soft excess. The coronal emission is described by \texttt{nthcomp} with the seed photons originating from the disc. In addition, we introduce \texttt{xillverCp} \citep{Garcia2010, Garcia2011,Garcia2013} to account for the reprocessed emission from the disc with reflection fraction $F_{\rm ref}$ fixed to $-1$.  In XSPEC notation, the phenomenological model reads as \texttt{tbnew} $\times$ \texttt{zmshift} $\times$ (\texttt{bbody}+\texttt{nthcomp}+\texttt{xillverCp}). In our spectral analysis, the fit is insensitive to the inclination angle, so we fix it to a best-fit value $i=73^{\circ}$. We also fix inclination angle at some other lower values and we get consistent results on the flux of \texttt{xillverCp}. In \texttt{nthcomp} model, we fix $kT_{\rm bb} = 10$~eV, which is the typical temperature for the accretion discs of AGNs and insensitive to the X-ray fitting processes \citep{Done2012}. The electron temperature $kT_{\rm e}$ and spectral slope $\Gamma$ are linked to that in \texttt{xillverCp}. The best-fit results are shown in the third column of Tab.~\ref{best-fit-1} and the uppermost panel of Fig.~\ref{ratio_chi} shows the corresponding data-to-model ratio and the zoomed-in version of the residuals in the 6~keV region.

Model~1 shows that the phenomenological blackbody model can fit the spectra well from a statistical point of view, with a good fit statistic $\chi_{\rm red}^{2} \sim 1.106$. The fitting requires a Galactic column density value of $N_{\rm H}$ = $0.041 \times 10^{22}$~cm$^{-2}$, which is consistent with the results in \citet{Dickey1990} and \citet{Willingale2013}. We set the ionization parameter as free and we get a value $\log\xi < 0.5$, hitting the lower limit $\log \xi=0.0$. From the upper right panel of Fig.~\ref{ratio_chi}, we can conclude that model 1 describes the iron emission region well with a neutral distant reflection model, although we can still see some systematic residuals around the Fe K energies, i.e. around 8~keV, which is the blue wing of a potential broad iron line (can be confirmed with model~3). The best-fit model gives a blackbody temperature $kT_{\rm bb}=0.143$~keV, which is in the range of characteristic temperature over a wide range of AGN luminosities and black hole masses (e.g., \citealt{Walter1993}; \citealt{Gierlinski2004}; \citealt{Bianchi2009}; \citealt{Crummy2006}). {The observed range in its temperature is very small, despite large changes in black hole mass, luminosity and spectral index across the AGN sample. Besides, this temperature ($\sim 0.143$~keV) is too hot for the disc at a sub-Eddington rate ($\lambda_{\rm Edd} = 0.2\%$, see Sec.~\ref{discussion} for detail) \citep{Gierlinski2004b}. Then, it favors an origin through atomic processes instead of purely continuum emission (e.g., \citealt{Gierlinski2004}; \citealt{Crummy2006}).}

The most peculiar aspect of this model is the relatively low temperature of the hot corona ( $kT_{\rm e} = 13.2_{-1.7}^{+2.5}$~keV), which is uncommon for AGNs (e.g., \citealt{Nandra1994}; \citealt{Ricci2017}; \citealt{Balokovic2020}; \citealt{Kang2022}; \citealt{Kamraj2022}). To check the potential degeneracy between the coronal temperature and the strength of reflection, we test the constraints on corona temperature (${kT}_{\rm e}$) and reflection fraction, presented in the upper panel of Fig.~\ref{contour}. Note that reflection fraction represents the so-called observer's reflection fraction \citep{Ingram2019} in our analysis, and is defined as the observed reflected flux divided by the observed hot coronal flux in the 0.3$-$80~keV band. From the contour plot, we find both parameters are tightly constrained, and there is not any degeneracy between two parameters. We now implement some more physically motivated model to study the soft excess and the excess around 6-7~keV.

\subsection{Model 2: warm corona model}

A warm ($T_{\rm e} \sim 10^{5-6}$~K) and optically thick ($\tau \sim $10–40) corona model has been proposed to explain the observed soft excess in AGNs (e.g., \citealt{Magdziarz1998}; \citealt{Petrucci2018}; \citealt{Porquet2018}; \citealt{Middei2020}). In this scenario, the soft excess is the high-energy tail of the resulting spectrum of a warm corona. This corona may be an extended, slab-like plasma at the upper layer of the disc, which is cooler than the hot ($T_{\rm e} \sim 10^{8-9}$~K), centrally located, and more compact corona responsible for the non-thermal power-law continuum.

Based on model~1, in model~2 we replace \texttt{bbody} with the physically motivated model \texttt{nthcomp} to represent the warm corona. In this case, the model reads as \texttt{tbnew} $\times$ \texttt{zmshift} $\times$ (\texttt{nthcomp1}+\texttt{nthcomp2}+\texttt{xillverCp}) in XSPEC language. \texttt{nthcomp1} is to model the soft excess and \texttt{nthcomp2} is to model the hot corona. To model  the reflection spectrum, we still use \texttt{xillverCp} with $F_{\rm ref}$ fixed to $-1$. \texttt{nthcomp} is characterized by the continuum slope, $\Gamma$, the temperature of the covering electron gas, $kT_{\rm e}$, and the seed photon temperature, $kT_{\rm bb}$. We fix $kT_{\rm bb} = 10$~eV for \texttt{nthcomp1} and \texttt{nthcomp2}. The electron temperature $kT_{\rm e}$ and spectral slope $\Gamma$ are set to be variable in \texttt{nthcomp1}. And the electron temperature $kT_{\rm e}$ and spectral slope $\Gamma$ of \texttt{nthcomp2} are linked to that in \texttt{xillverCp}.

\begin{figure*}
    \centering
    \includegraphics[width=0.85\linewidth]{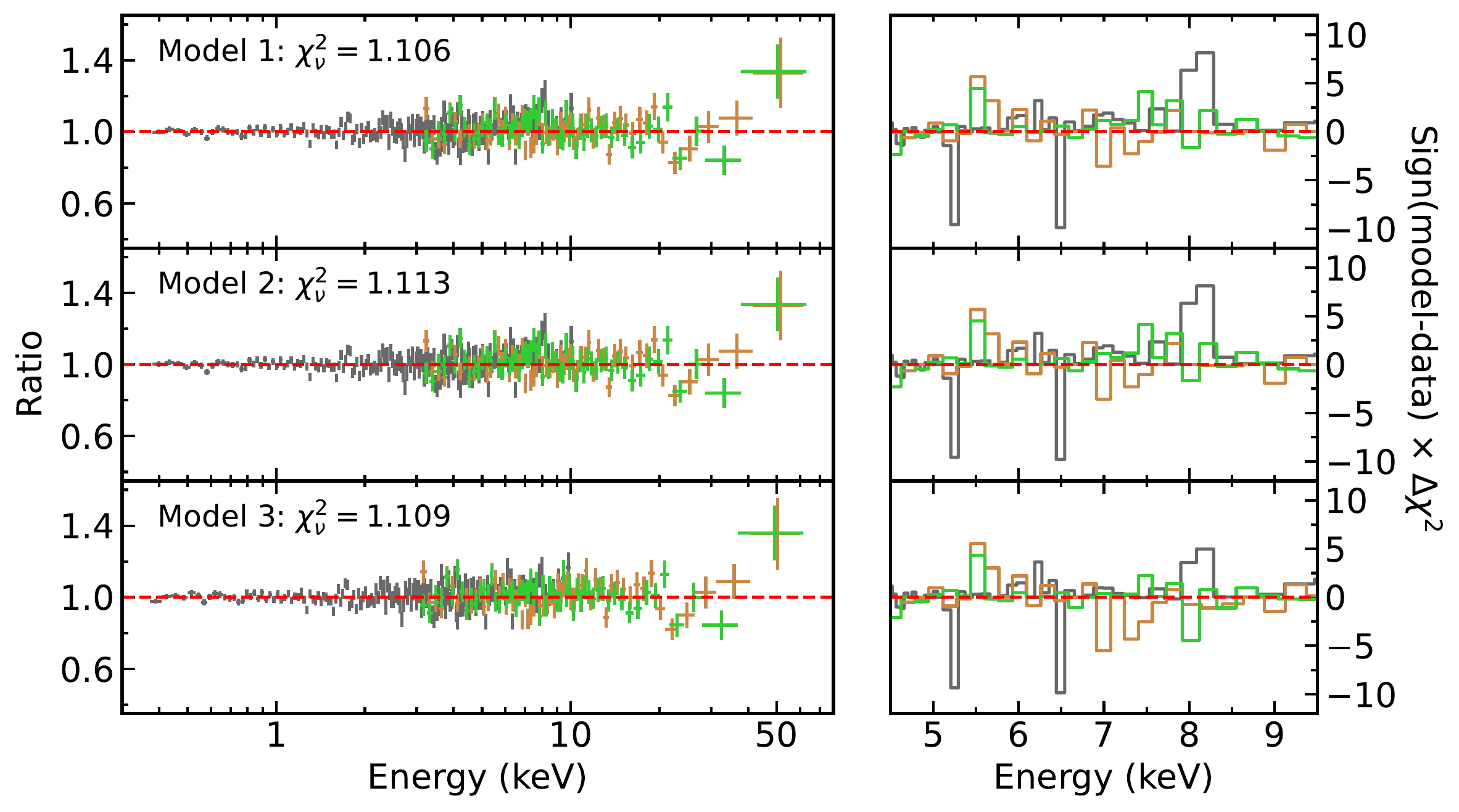} \\
    \caption{Left: Data-to-model ratios for the blackbody model (model~1, top), the warm corona model (model~2, middle), and the relativistic reflection model (model~3, bottom). The \xmm\, \nustar/FPMA data and \nustar/FPMB are marked with gray, orange and green crosses respectively. Right: The zoom-in residuals between 4.5–9.5 keV. The best-fit models are shown in Fig.~\ref{eemod}.}
    \label{ratio_chi}
\end{figure*}

The warm corona model results in a fit statistic $\chi_{\rm red}^{2} \sim 1.113$ (see the fourth column of Tab.~\ref{best-fit-1}) and models the residuals in Fig.~\ref{soft_excess_iron} well (the middle panel of Fig.~\ref{ratio_chi}). The spectral slope found for the warm corona is $\Gamma = 2.6_{-0.4}^{+0.4}$ and the estimate of the temperature is $0.1908_{-0.04}^{+0.0009}$, which is consistent with the range ($0.1 \sim 1$ keV) reported in \citet{Petrucci2018}. By comparison, the hot corona is characterized by a more gentle spectral slope ($\Gamma=1.716_{-0.009}^{+0.009}$) and a higher temperature (${kT}_{\rm e}=13.23_{-0.9}^{+0.21}$~keV), which are almost identical to model 1. The middle panel of Fig.~\ref{contour} shows the constraints on the temperature of the hot corona (${kT}_{\rm e}$) and the reflection fraction for model~2.

\begin{table*}
    \centering
    \caption{Best-fit values for model~1, model~2 and model~3.} \label{best-fit-1}
    \renewcommand\arraystretch{1.5}
    \setlength{\tabcolsep}{6mm}
    \begin{tabular}{lcccc}
    \hline\hline
    & & Model~1 & Model~2 & Model~3 \\
    \hline
    Model & Parameter & & & \\
    \texttt{tbnew} & $N_{\rm H}$ (10$^{22}$~cm$^{-2}$) & $0.041_{-0.007}^{+0.009}$  & $0.061_{-0.013}^{+0.003}$ & $0.034_{-0.004}^{+0.005}$ \\
    \hline
    \texttt{bbody} & $kT$ (keV) & $0.143_{-0.011}^{+0.017}$ & - & - \\
    \texttt{nthcomp} & $\Gamma$ & - & $2.6_{-0.4}^{+0.4}$ & - \\
                     & $kT_{\rm e}$ (keV) & - & $0.1908_{-0.04}^{+0.0009}$ & - \\
    
    \texttt{relconv} & $a^*$ & - & - & $0.998^*$ \\
    \texttt{xillverDCp} & $\log n_{\rm e}$ & - & - & $18.1_{-1.0}^{+0.4}$ \\
                          & $\log \xi$ & - & -  & $1.0_{-0.7}^{+0.6}$ \\
    \hline
    \texttt{nthcomp} & $\Gamma$ & $1.716_{-0.009}^{+0.01}$ & $1.716_{-0.009}^{+0.009}$ & $1.753_{-0.02}^{+0.014}$ \\
                    & $kT_{\rm e}$ (keV) & $13.2_{-1.7}^{+2.5}$ & $13.23_{-0.9}^{+0.21}$ & $13.8_{-1.9}^{+2.5}$\\
    \hline
    \texttt{xillverCp} & $A_{\rm Fe}$ & $10.0_{-1.5}^{+P}$ & $10.0_{-0.9}^{+P}$ & $8.1_{-3}^{+P}$ \\
                    & $i$ (deg) & $73^*$ & $73_{-15}^{+7}$ & $72.6_{-5}^{+4}$ \\
                    & $\log \xi$ & $0.0_{-P}^{+0.5}$ & $0.0_{-P}^{+0.4}$ & $0^*$ \\
    
    \hline
    & $C_{\rm FPMA}$ & $1.247_{-0.022}^{+0.023}$ & $1.247_{-0.016}^{+0.016}$ & $1.248_{-0.022}^{+0.022}$ \\    
    & $C_{\rm FPMB}$ & $1.226_{-0.022}^{+0.023}$ & $1.226_{-0.016}^{+0.016}$ & $1.228_{-0.022}^{+0.022}$ \\
    \hline
    & $F_{\rm bbody}$ ($\times$ 10$^{-13}$) & $2.6_{-0.9}^{+1.0}$ & - & - \\
    & $F_{\rm WC}$ ($\times$ 10$^{-13}$) & - & $4.7_{-1.8}^{+1.3}$ & - \\
    & $F_{\rm xillverCp}$ ($\times$ 10$^{-13}$) & $5.9_{-0.8}^{+1.7}$ & $5.9_{-0.6}^{+0.6}$ & $6.2_{-1.5}^{+2.5}$ \\
    & $F_{\rm HC}$ ($\times$ 10$^{-11}$) & $1.55_{-0.03}^{+0.06}$  & $1.51_{-0.04}^{+0.05}$ & $1.41_{-0.06}^{+0.06}$ \\
    & $F_{\rm RR}$ ($\times$ 10$^{-12}$) & -  & - & $1.2_{-0.3}^{+0.3}$ \\
    \hline
    $\chi^2$/d.o.f &  & 646.39/585 & 649.51/584 & 648.11/584 \\
    \hline\hline 
    \end{tabular} \\
    \vspace{0.2cm}
    \textit{Note.} The flux (0.3--80 keV) of the each component are presented in units of erg~s$^{-1}$~cm$^{-2}$. $F_{\rm WC}$, $F_{\rm HC}$ and $F_{\rm RR}$ represent the flux of the warm corona component, the hot corona component and the relativistic reflection component, respectively. $\xi$in units of erg~cm~s$^{-1}$. $n_{\rm e}$ in units of cm$^{-3}$.
\end{table*}



\subsection{Model 3: relativistic reflection model}

The other popular explanation of the soft excess is the relativistic disc reflection model (e.g., \citealt{Crummy2006}; \citealt{Fabian2009}; \citealt{Walton2013}; \citealt{Jiang2018}). In the strong gravitational field of supermassive black holes, the fluorescent features radiated on the inner region of the accretion disc are blurred \citep{Fabian2005}. Moreover, it has recently been shown that the existence of the enhanced inner-disc density, above the commonly assumed value of $n_{\rm e} = 10^{15}$~cm$^{-3}$ (e.g., \citealt{Ross1993}, \citealt{Ross2005}; \citealt{Garcia2011}), results in an increased emission at soft energies ($<$2 keV). This occurs because at high densities free–free heating (bremsstrahlung) in the disc atmosphere becomes dominant and results in an increased gas temperature (\citealt{Garcia2016}; \citealt{Jiang2019c}). As a consequence, relativistic reflection from a highly dense disc can lead to increased low-energy emission, which may account for the soft excess.

To test this explanation, We implement the reflection model \texttt{xillverDCp} \footnote{https://sites.srl.caltech.edu/~javier/xillver/index.html}, a version of \texttt{xillver} that allows for variable disc density, convolved by \texttt{relconv} \citep{Dauser2013} to model the relativistic reflection component. And the bare \texttt{xillverCp}, in which electron density is fixed at $n_{\rm e} = 10^{15}$~cm$^{-3}$, is reserved for the distant non-relativistic reflection component. From the view of self-consistent model, the parameters $F_{\rm ref}$ of the two reflection models are fixed as $-1$ to return only the reflection component, and a non-thermal power-law continuum model \texttt{nthcomp} is included to account for the power-law component from the hot corona. In XSPEC language, the model combination is \texttt{tbnew} $\times$ \texttt{zmshift} $\times$ (\texttt{relconv} $\times$ \texttt{xillverDCp} + \texttt{nthcomp} + \texttt{xillverCp}). Same as model~2, we fix $kT_{\rm bb} = 10$~eV for \texttt{nthcomp}. For the spectral slope $\Gamma$ and electron temperature ${kT}_{\rm e}$, the hot corona, relativistic disc reflection, and distant reflection components are tied together. And for \texttt{xillverCp} and \texttt{xillverDCp}, the inclination angles are linked to the same parameter in \texttt{relconv}. The ionization parameter is set to its minimum ($\xi = 0$~erg~cm~s$^{-1}$) in \texttt{xillverCp} and free in \texttt{xillverDCp}. Similarly, the electron density is set to its minimum ($n_{\rm e} = 10^{15}$~cm$^{-3}$) in \texttt{xillverCp} and free in \texttt{xillverDCp}. The inner radius $R_{\rm in}$ and the outer disc radius $R_{\rm out}$ of the accretion disc are fixed at their default value, i.e., $R_{\rm in}=R_{\rm ISCO}$ and $R_{\rm out}=400R_{\rm g}$ ($R_{\rm g}=GM/c^2$, gravitational radii). We assume that the emissivity profile follows a $q=3$ power law and the spin parameter is fixed at the maximum value $a_*=0.998$ in \texttt{relconv} because of its insensitivity to the fit, and the disc inclination angle varies freely. The  iron abundance of \texttt{xillverCp} is linked to that in \texttt{xillverDCp}.

\begin{figure}
    \centering
    \includegraphics[width=0.85\linewidth]{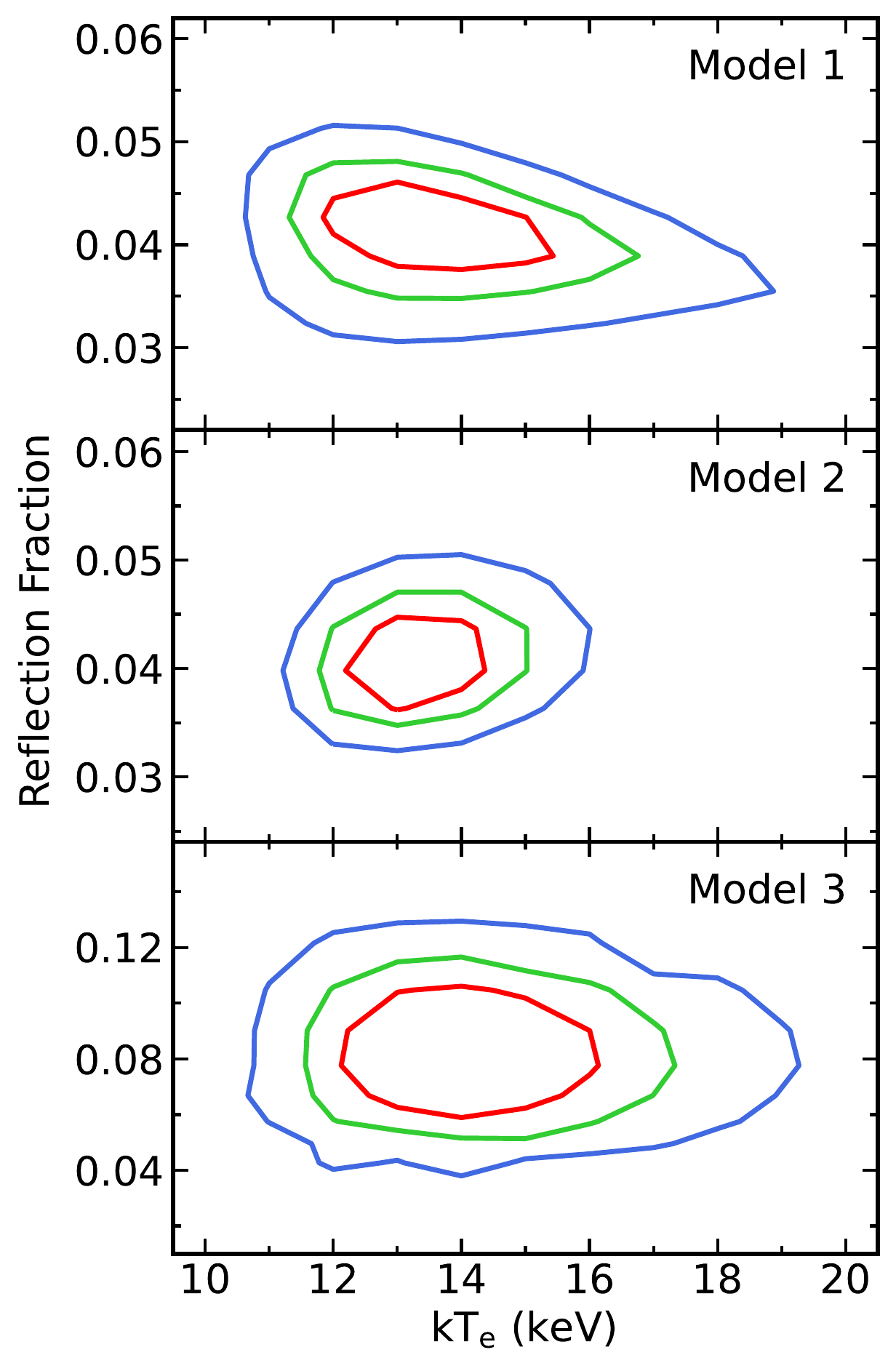} \\
    \caption{Constraints on temperature of hot corona (${kT}_{\rm e}$) and . For model~3, the reflection fraction is defined with the method in text. The constraints on ${kT}_{\rm e}$ are consistent from the various model. The red, green, and blue curves represent, respectively, the 68\%, 90\%, and 99\% confidence level limits for two relevant parameters ($\Delta\chi^2 = 2.30$, 4.61, and 9.21, respectively).}
    \label{contour}
\end{figure}

\begin{figure}
    \centering
    \includegraphics[width=0.85\linewidth]{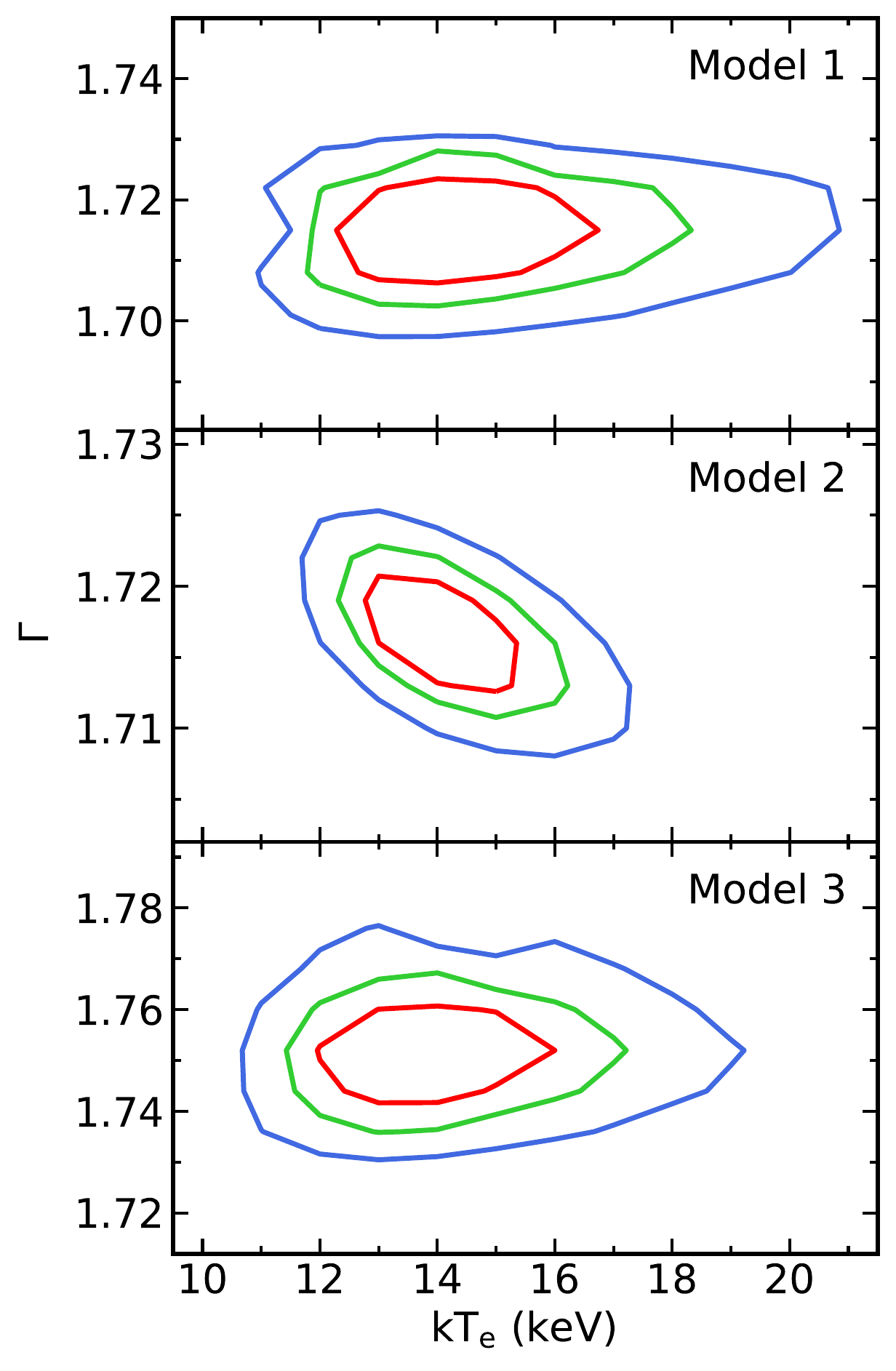} \\
    \caption{Constraints on temperature of hot corona (${kT}_{\rm e}$) and photo index ($\Gamma$). The red, green, and blue curves represent, respectively, the 68\%, 90\%, and 99\% confidence level limits for two relevant parameters ($\Delta\chi^2 = 2.30$, 4.61, and 9.21, respectively).}
    \label{contour_gamma_kT}
\end{figure}

\begin{figure}
    \centering
    \includegraphics[width=0.85\linewidth]{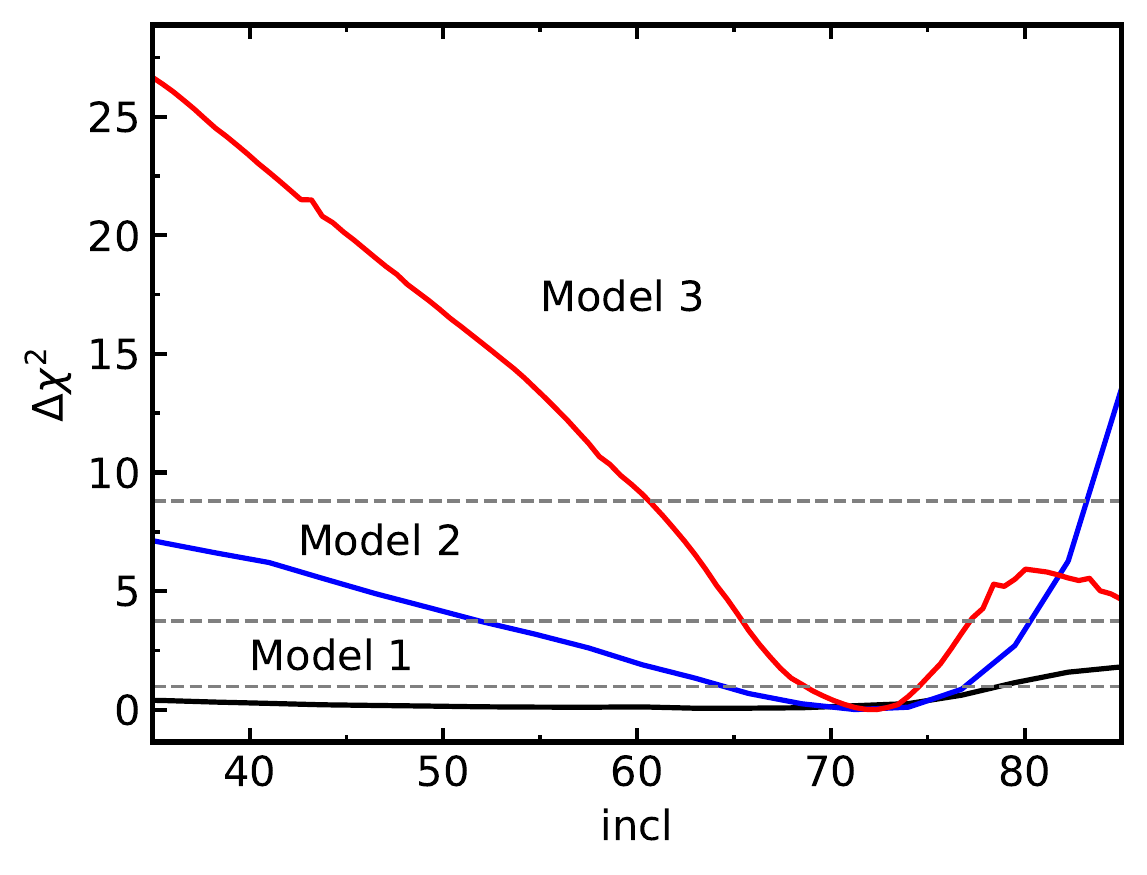} \\
    \caption{{The $\Delta\chi^2$ contours for the inclination parameters for all 3 models. The gray horizontal lines represent the $1\sigma$, $2\sigma$ and $3\sigma$ confidence levels from the bottom up for a single parameter of interest.}}
    \label{incl_steppar}
\end{figure}

\begin{figure}
    \centering
    \includegraphics[width=0.9\linewidth]{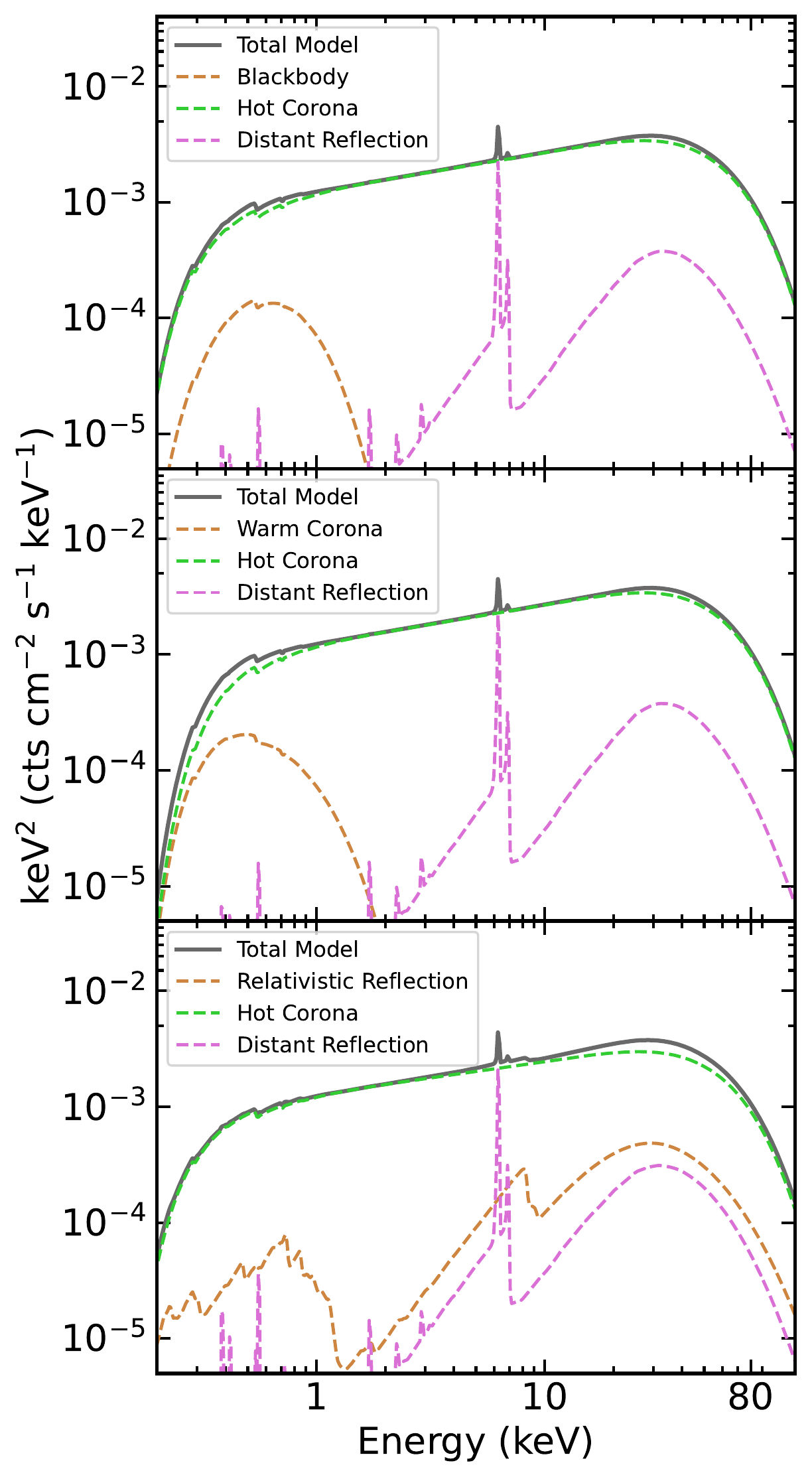} \\
    \caption{Model components of the best-fit model in the blackbody model (model~1, top), the warm corona model (model~2, middle), and the relativistic reflection model (model~3, bottom). }
    \label{eemod}
\end{figure}

The best-fit values are listed in the fifth column of Tab.~\ref{best-fit-1} and the residuals are shown in the lower panel of Fig.~\ref{ratio_chi}. The relativistic reflection picture provides a better fit than the warm corona model, with $\chi_{\rm red}^{2} \sim 1.110$. Compared with the warm corona model, the relativistic reflection model slightly improves the fit with $\Delta \chi^2 = 1.95$. As shown in the fifth column of Tab.~\ref{best-fit-1}, this model gives essentially similar results to those reported in model~1 and model~2. With such a relativistic reflection model, the iron complex region is modeled better as shown in the lower right panel of Fig.~\ref{ratio_chi}. Same as the previous 2 models, this model provides a low temperature for the hot corona as well. The constraints on the temperature of the hot corona (${kT}_{\rm e}$) and reflection fraction are presented in the lower panel of Fig.~\ref{contour}. Here, the reflection fraction is flux ratio between the relativistic reflection component (dominates in 0.3$-$80~keV band) and the hot corona component. {We test the constraints on corona temperature ($kT_{\rm e}$) and photo index ($\Gamma$) to check the potential degeneracy in this parameter plane. The results are presented in Fig.~\ref{contour_gamma_kT}. It shows that there is not significant degeneracy between $kT_{\rm e}$ and $\Gamma$ in all models. We also find that a high inclination is preferred by 3 models, though it can not be constrained well in model~1 (Fig.~\ref{incl_steppar}). }

In the fit of model~3, we model the emissivity profile of the disc with a power-law mode with $q_{\rm in}=q_{\rm out}=3$. If we free $q_{\rm in}$ and $q_{\rm out}$ in the fit, we find that these two parameters are insensitive to the fit. So we do not explore further such a possibility.


\section{Discussion}
\label{discussion}

In the previous sections, we presented the spectral properties of the AGN \src, and found that the variability mainly happens over the entire broadband spectrum, and the hardness ratio curve remains constant. With a simple absorbed power-law model, the typical AGN spectral features---a soft excess below $\sim$2 keV, Fe K$\alpha$ emission at $\sim$6.4 keV, and a cutoff at high energy, are revealed. The soft excess can be modeled by a single temperature blackbody model \texttt{bbody} with a typical temperature $kT_{\rm bb} = 0.143$~keV, and iron emission can be modeled by a simple gaussian model with centroid energy $E = 6.39$~keV in the rest frame of the object and line width $\sigma = 0.04_{-0.03}^{+0.05}$~keV, based on which we identify this line as Fe K$\alpha$ emission line and an origin of a distant reflector.

We carry out spectral analyses based on two different hypotheses to explain the soft excess: the warm corona and the relativistic reflection scenario. In the process of fitting with the warm corona model, the introduction of a soft ($\Gamma = 2.6$), cool temperature ($kT_{\rm e} = 0.1908$~keV) model \texttt{nthcomp} yields good results. In the process of fitting with the relativistic reflection model, we utilize the xillver-based model and this model provides somewhat satisfactory fits to the spectra. In this section, we discuss the physical aspects of previous fits and the exploration of further study.

\subsection{Eddington ratio estimation}

We calculate the Eddington ratio $\lambda_{\rm Edd}$ of \src\ by applying an average bolometric luminosity correction factor $\kappa$ = 20 \citep{Vasudevan2007} to the 2–10 keV band unabsorbed luminosity $5.56 \times 10^{42}$~erg~s$^{-1}$. A black hole mass of $\sim 4.57 \times 10^8 M_{\odot}$ \citep{Ponti2012} is considered. We obtain $\lambda_{\rm Edd} = \kappa L_{\rm X}/L_{\rm Edd} = 0.002$, which is consistent with the results reported in \citet{Ghosh2021}. \citet{Ghosh2021} analyzed  the broadband spectra observed by \xmm\ and \textit{Suzaku}, and reported that the source was accreting at a sub-Eddington rate ($\lambda_{\rm Edd}$ varies within 0.002 - 0.008) between 2007 and 2012. They also reported a power-law continuum with a photon index varying between $\Gamma=1.7 - 2.0$, and the presence of a broad Fe emission line at $\sim$6.4 keV in the source spectra with $\sigma=0.08 - 0.14$~keV. It seems that the profile of Fe emission line is related to the flux state of the source. We will conduct a detailed investigation of the variability and spectral properties of different flux states in a forthcoming paper.


\begin{figure*}
    \centering
    \includegraphics[width=0.45\linewidth]{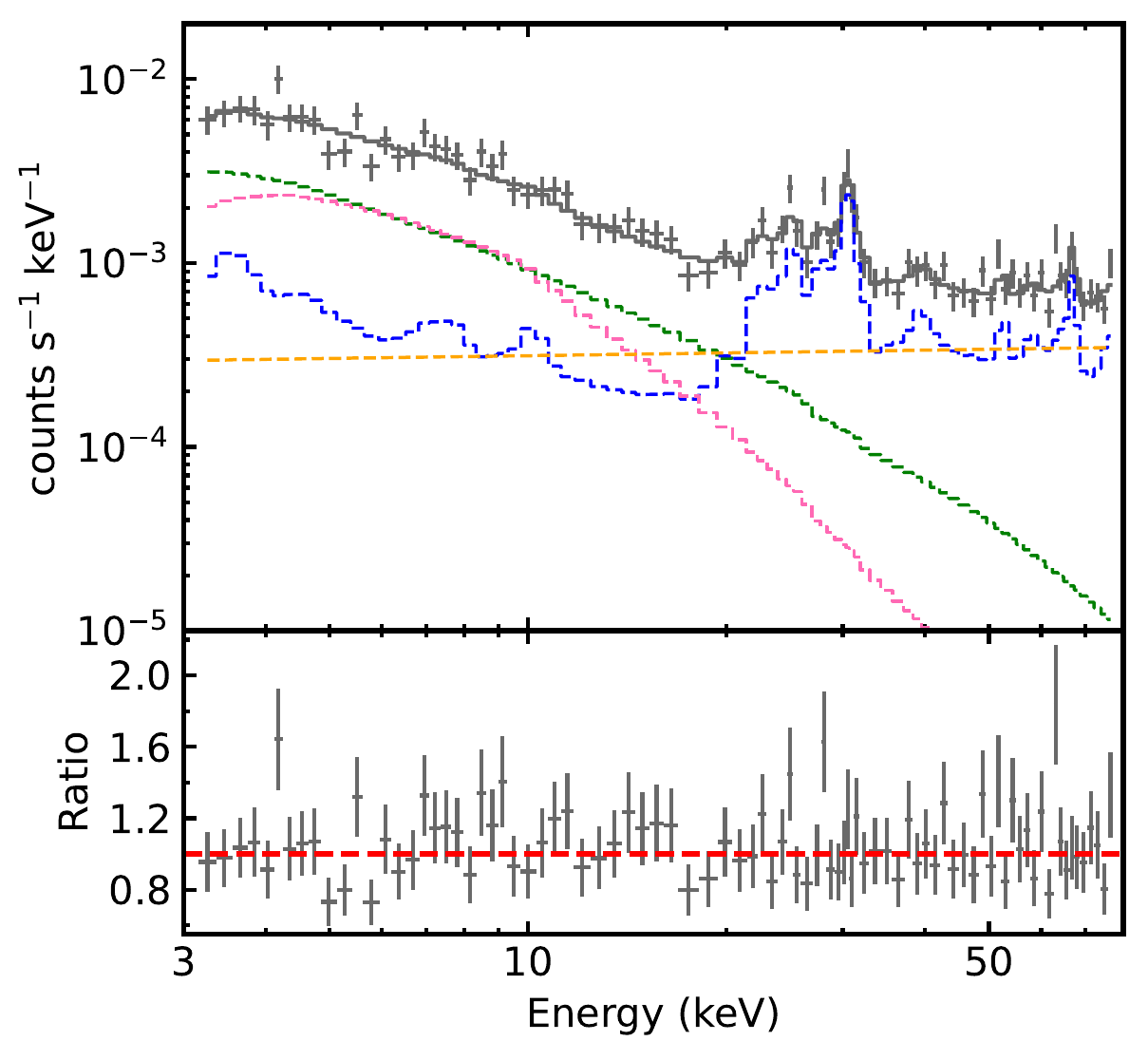}
    \includegraphics[width=0.45\linewidth]{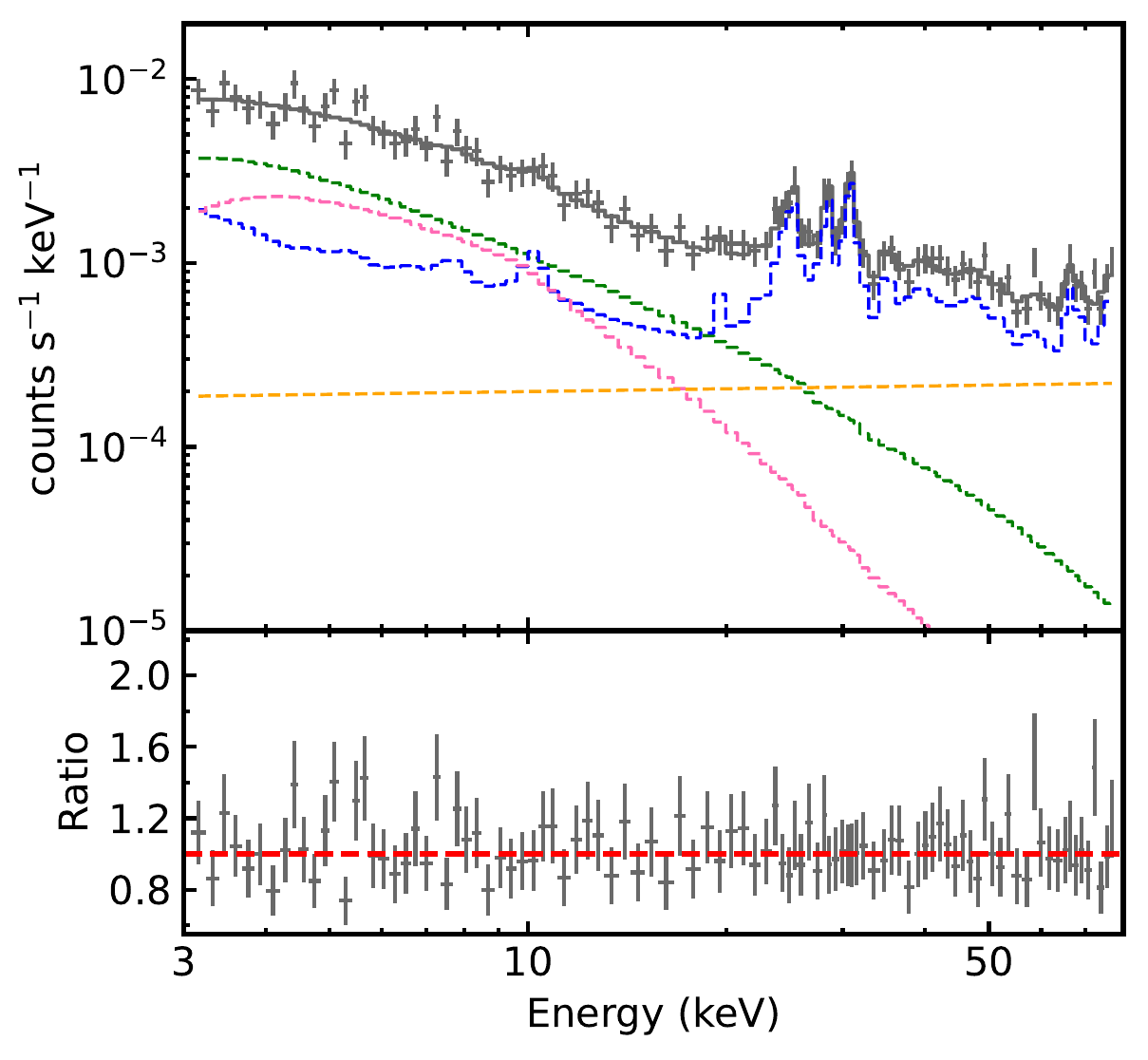}\\
    \caption{Left: the background spectrum of \nustar\ FPMA ( Obs ID: 60502035002) with \texttt{nuskybgd} model. The black, green, pink, blue, and orange lines, respectively, indicate the total background, aCXB, fCXB, Inst, and Intn, included in nuskybgd. Right: the background spectrum of \nustar\ FPMB of the same observation. Each component is shown with same color as left.}
    \label{back}
\end{figure*}


\begin{figure}
    \centering
    \includegraphics[width=0.85\linewidth]{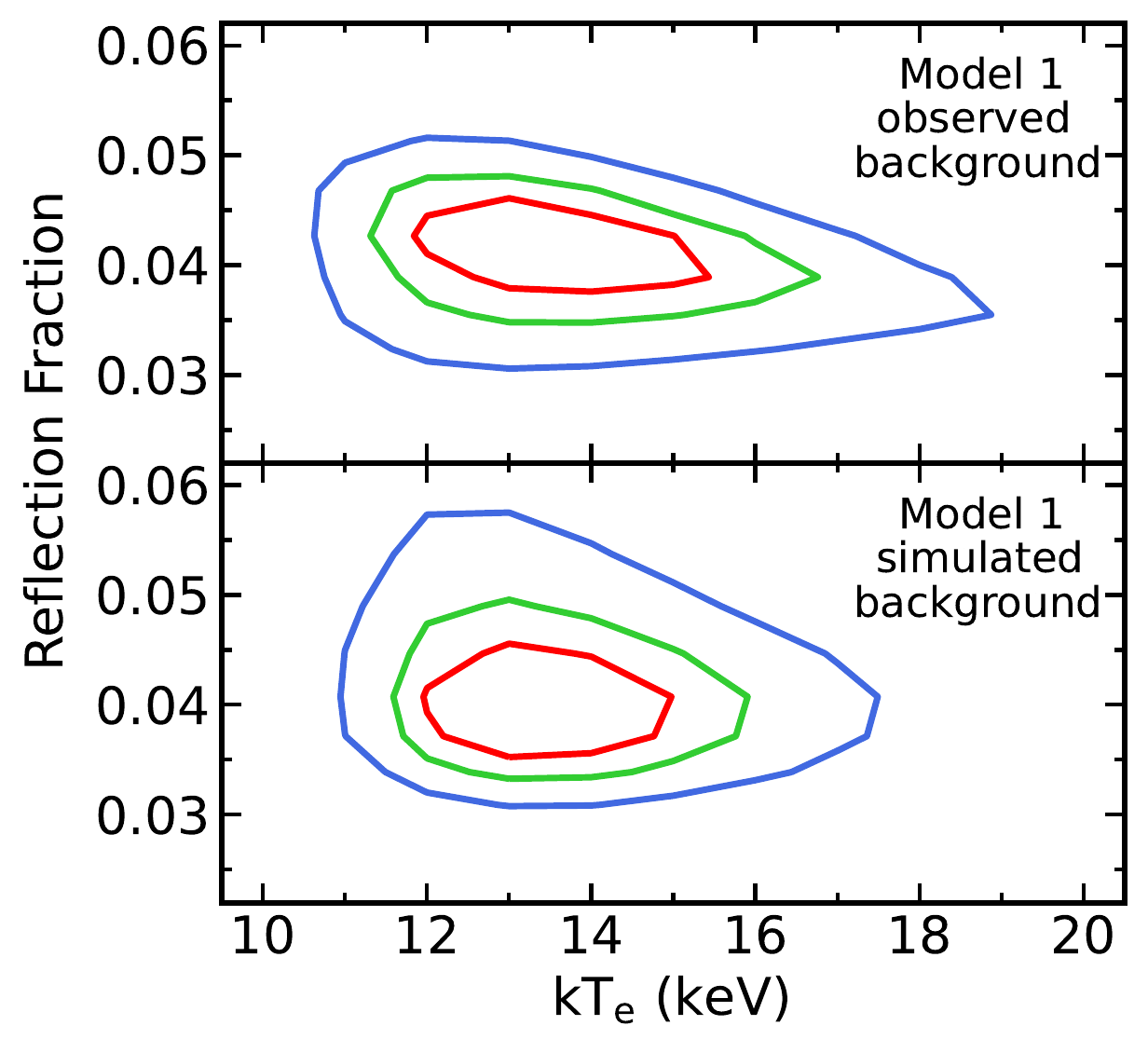}\\
    \caption{Constraints on temperature of hot corona and reflection fraction with model~1. Upper panel show the results using observated background. Lower panel show the results using simmulated background. The red, green, and blue curves represent, respectively, the 68\%, 90\%, and 99\% confidence level limits for two relevant parameters ($\Delta\chi^2 = 2.30$, 4.61, and 9.21, respectively).}
    \label{back_contour}
\end{figure}

\subsection{Model the background spectra}

As shown in Fig.~\ref{soft_excess_iron}, the hard X-ray spectral shape of \src\ depends on the accuracy of the background modelling on \nustar. To avoid the influence of uncertainty from the background, we carefully choose the background region and use the most strict filtering criteria to filter the data. Moreover, we conduct a detailed analysis of background spectra and the uncertainty from the background spectrum. To do so, we use the toolkit named “\texttt{nuskybgd}” \citep{Wik2014}. Thanks to \texttt{nuskybgd}, we can construct the background spectra for any region in which we are interested. 

As is typical, the background has both intrinsic and extrinsic components. The \texttt{nuskybgd} model consists of four components, which combine to fully describe the background: $$B_{\rm d}(E,x,y) = A_{\rm d}(E,x,y) + f_{\rm d}(E,x,y) + S_{\rm d}(E) + I_{\rm d}(E)$$ 
$A_{\rm d}(E,x,y)$ describes the stray-light cosmic X-ray background (CXB) through the aperture, marked by "aCXB"; $f_{\rm d}(E,x,y)$ describes the focused CXB, marked by "fCXB"; $S_{\rm d}(E)$ describes instrument line emissions and reflected solar X-rays, marked by "Inst"; $I_{\rm d}(E)$ describes the instrument Compton scattered continuum emissions, marked by "Intn". Using the ftool included in \texttt{nuskybgd}, we can fit the background spectrum by the model above. Based on the best-fit parameters, the background image and the background spectrum for an arbitrary region in the FOV can be produced.

Our background estimation of \src\ with \nustar\ is as follows. We extract the FPMA and FPMB background spectra for Epoch~1-5 separately, according to the strategy in Sec.~\ref{observations}. And then fit the background spectra with the standard \texttt{nuskybgd} model. The spectra of Epoch~1 are shown with all the components of the background model in Fig.~\ref{back}. The background spectra are well-fitted by our background model. To construct the background spectrum for the region of interest, we use the ftool \texttt{nuskybgd-spec} in \texttt{nuskybgd} for aCXB, fCXB, Inst, and Intn components. The background is estimated based on the best-fit parameters of the background spectrum for the region of interest. Thus, we simulate FPMA and FPMB background spectra for every epoch, which is 10 times the exposure time of the original observation. Because the model for background spectra is complex, the error would be very large if the exposure is too short. Simulated background spectra are merged to an averaged spectra for FPMA and FPMB separately. We replace the background spectra with the simulated one in model 1 to estimate the uncertainty from the background. Fig.~\ref{back_contour} compares the measurements using two methods. An almost identical constraint is given with the simulated background.

As shown in Fig.~\ref{back}, at higher ($E > 30$~keV) energies, internal $I_{\rm d}(E)$ term strongly dominates the background spectra. The remainder of the internal background consists of various activation and fluorescence lines, which are mostly resolved and only dominate the background between 22–32 keV. Above these energies, weaker lines are still present, but the continuum dominates. There is no dependence on pixel location $x$, $y$, only on energy $E$, so the spatial distribution across a given detector is uniform. and for internal background, the systematic uncertainty could in theory be arbitrarily close to 0\% \citep{Wik2014,Tsuji2019}, or at least has the least uncertainty compared to other components. In summary, \nustar\ observation supply the relatively reliable broadband spectra from 3.0-78.0~keV, although the background plays an important role in the high energy band.

\subsection{Physical properties of the warm corona model}

In the fit of model~2, the warm corona model with a hot corona and a neutral distant reflection component describes the observational data well. The corresponding optical depth of the warm corona ($\tau \sim 18-25$), calculated with Eq. (13) in \citet{Beloborodov1999}, i.e., $\Gamma \simeq \frac{4}{9} y^{-\frac{2}{9}}$ with $y =4[kT_{e}/m_{e}c^2+(kT_{e}/m_{e}c^2)^2]\tau(\tau + 1)$ the so-called Compton parameter. \citet{Petrucci2018} test the warm corona model on a statistically significant sample of unabsorbed, radio-quiet AGNs with \xmm\ archival data and find the temperature of the warm corona to be uniformly distributed in the 0.1–1 keV range, while the optical depth is in the range $\sim$ 10–40. The observational characteristics of the warm corona (i.e., a photon index of 2.5 and a temperature of 0.1–2~keV) is within the prediction of \citet{Petrucci2018} ($\tau \sim$10–40), and agree with an extended warm corona covering the disc which is mainly nondissipative (\citealt{Petrucci2018}, later corrected in \citealt{Petrucci2020}).

With the warm corona scenario, we get a slightly harder slope of the continuum $\Gamma= 1.716$, compared with model~3. Similar results were also be found in \citet{Garcia2018}, and \citet{Xu2021}. It means that the warm corona scenario requires a harder continuum in the absence of the compensation of the high energy photons from the inner disc reflection. In our fits, the difference in the photon index between the warm corona and relativistic reflection is $\Delta \Gamma \sim 0.04$. For the given X-ray AGN spectrum data with the soft excess, the lack of any disc reflection component in the pure warm corona model is likely to provide a harder continuum than the relativistic reflection explanation.

However, \citet{Gronkiewicz2020} computed the transition from the disc to corona, using the vertical model of the disc supported and heated by the magnetic field together with radiative transfer in hydrostatic and radiative equilibrium. They concluded that the radial extent of the warm corona is limited by local thermal instability and a warm corona like this is stronger in the case of a higher accretion rate and a greater magnetic field strength. So thermal instability should prevent the warm corona from forming for lower accretion rate system. The low accretion rate system is unable to provide enough energy to sustain a warm corona \citep{Ballantyne2020}. It is therefore unclear whether a strong warm corona can be sustained at the low accretion rates relevant here ($\dot{m}\sim0.002$). This may imply that even if a warm corona is present, a contribution from the disc reflection would be necessary to produce the observed strong soft excess.

\subsection{Physical properties of the relativistic disc reflection}

The high-density disc reflection model proposed in \citet{Garcia2016} is based on an extended model of the standard accretion disc. \citet{Garcia2016} demonstrated that if the disc density is higher than the typically fixed value $n_{\rm e} = 10^{15}$~cm$^{-3}$, the main effect is the enhancement of the reflected continuum at low energies, further enhancing the soft excess. We note that the relativistic reflection model produces a similar statistical result to the warm corona model for \src, with consistent key parameters. 

The relativistic reflection explanation requires a dense accretion disc with density $\log [n_{\rm e}/$cm$^{-3}]=18.1_{-1.0}^{+0.4}$. This result is consistent with previous findings that a larger gas density than the previously adopted value of $\log [n_{\rm e}/$cm$^{-3}]=15$ is usually required for SMBHs with $\log [m_{\rm BH}/M_{\odot}] \le 8$, like Ark~564 \citep{Jiang2019c} and ESO 362$-$G18 \citep{Xu2021}. Another factor that affects the expected disc density is the accretion rate. {According to the standard thin disc model \citep{Shakura1973}, \citet{Svensson1994} derived a relationship between the density $n_{\rm e}$  of a radiation-pressure-dominated disc and the accretion rate $\Dot{m}$, $$n_{\rm e}=\frac{1}{\sigma_{\rm T}R_{\rm S}}\frac{256\sqrt{2}}{27}\alpha^{-1} R^{3/2}\Dot{m}^{-2}[1-(\frac{R_{in}}{R})^{1/2}]^{-2}[\xi(1-f)]^{-3},$$ where $\sigma_{\rm T} = 6.64 \times 10^{25} cm^2$ is the Thomson cross-section; R is in the units of $R_{S}$ ( Schwarzschild radius); $\xi$ is the conversion factor in the radiative diffusion equation and chosen to be 1, 2, or 2/3 by different authors; $f$ is the fraction of the total transported accretion power released from disc to the hot corona.} With the correlation formula $\log[n_{\rm e}] \propto -2\log[\dot{m}]$, they concluded that a lower accretion rate leads to a higher gas density. Similar conclusions were found in disc reflection modelling of black hole (BH) X-ray binaries \citep{Jiang2019b} and other AGNs with a high BH mass \citep[e.g. Mrk~509,][]{Garcia2018}.

Another characteristic is the low ionization parameter of the relativistic reflection component, which indicates that the degree of ionization on the accretion disc is relatively low. \citet{Ballantyne2011} reported a positive statistical correlation between $\xi$ and the AGN Eddington ratio $\dot{m}$ based on the simple $\alpha$-disc theory. We plug the Eddington ratio $\dot{m}=0.002$ into Formula (1) of \citet{Ballantyne2011}, and get the estimation of the ionization parameter through, $\log\xi \sim 0.5$. This value is smaller than our fitting results, but consistent within error. The physical interpretation is that the accretion rate affects the fraction of the accretion energy dissipated in the corona (e.g., \citealt{Svensson1994}; \citealt{Merloni2002}; \citealt{Blackman2009}), which emits X-ray photons to photoionize the inner disc surface. All models show a low reflection fraction, which recall the case of an outflowing corona (\citealt{Beloborodov1999b}; \citealt{Malzac2001}). This can be confirmed by future missions.

We explore the possibility of measuring the spin with this model. The result is that the spin parameter cannot be constrained in \texttt{relconv}. With one more free parameter, we slightly improve the fit with $\Delta \chi^2= 0.95$ and only have a lower limit ($a_*>-0.58$).


\begin{figure}
    \centering
    \includegraphics[width=0.9\linewidth]{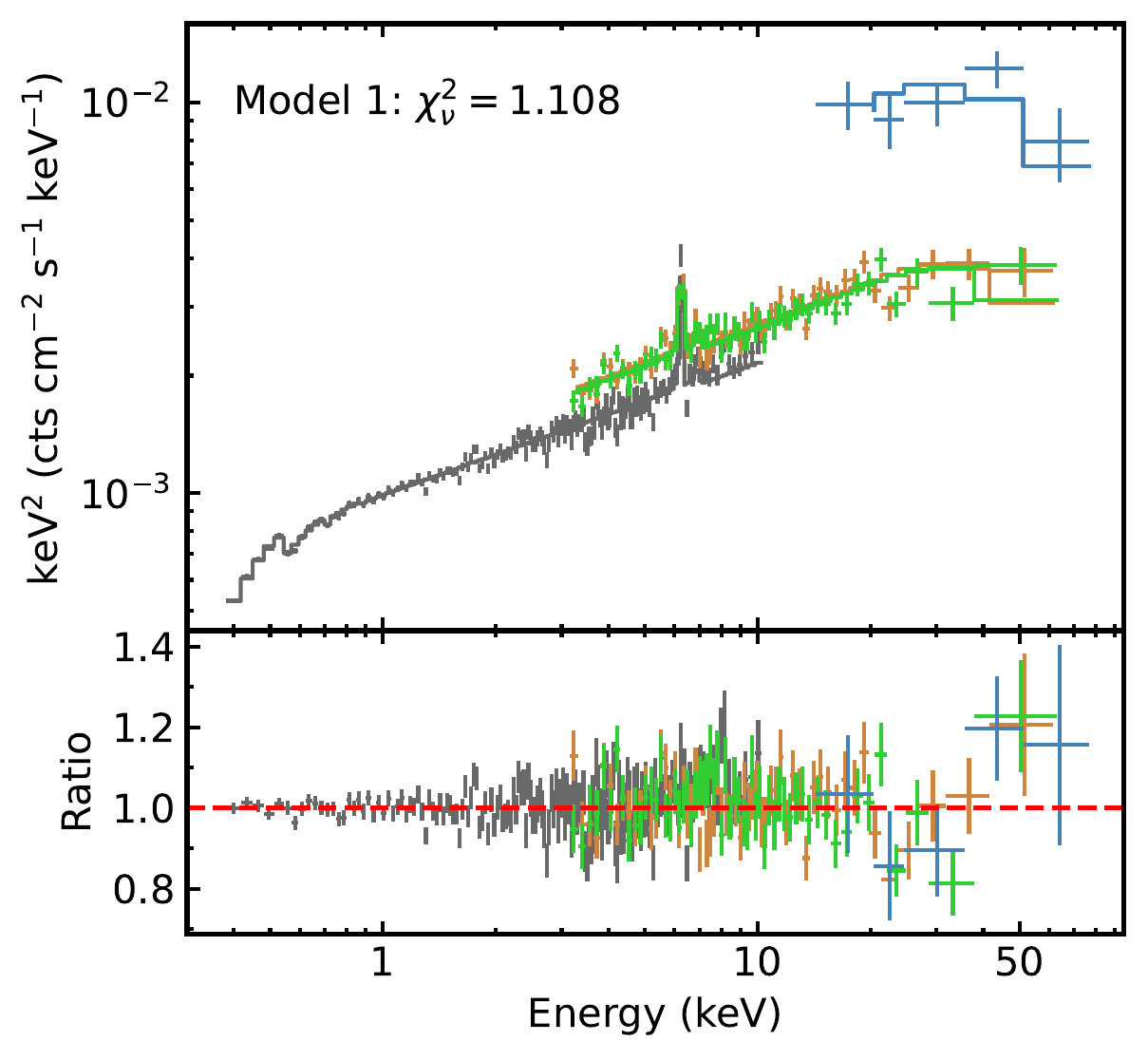} \\
    \caption{The spectrum and the corresponding residuals with model~1, considering \textit{Swift}/BAT spectum (blue). The \xmm\, \nustar/FPMA data and \nustar/FPMB are marked with gray, orange and green crosses respectively. }
    \label{swift_ratio}
\end{figure}

\subsection{Low-temperature corona in sub-Eddington accretors}

The most striking discovery is the relatively low temperature of the hot corona, which is confirmed by all broadband models, as shown in Fig.~\ref{contour}. To seek any possible variability of the temperature of coronae in a long timescale, we fit the \textit{Swift} 70-month BAT spectrum \citep{Baumgartner2013} together with the \xmm\ and \nustar\ spectrum. The ratio plot is shown in Fig.~\ref{swift_ratio}, fitted with model 1. The \textit{Swift} BAT spectrum shows a consistent shape with the \nustar\ FPM spectra above 30~keV, and we get an almost identical fitting results. The cross-calibration constant for BAT is 3.64.

\citet{Fabian2015} compiled a sample of all the high energy cut-offs observed with \nustar\ and populated these sources on the compactness-temperature ($\ell$–$\Theta$) plane, where $\Theta = kT_{\rm e}/m_{\rm e}c^{2}$ is the coronal electron temperature normalized by the electron rest energy and  $l= (L/R)(\sigma_{\rm T}/m_{\rm e}c^{3}$) is the dimensionless compactness parameter \citep{Guilbert1983}. \citet{Fabian2015} defined $L$ as the luminosity of the power-law component from 0.1–200 keV and $R$ as the radius of the corona (assumed spherical). 

The allowed parameter space in the $\ell$-$\Theta$ plane is limited by theoretical constraints, like the pair balance line that is estimated by \citet{Svensson1984}. With more power is fed into the corona, electron temperature $\Theta = kT_{\rm e}/m_{\rm e}c^{2}$ rises, and Compton scattering of the soft photons produces a power-law radiation spectrum extending to a Wien tail at energies around $2\Theta$. When the tail extends above $\sim2m_{\rm e}c^{2}$, photon–photon collisions happen and create electron–positron pairs. Thus any additional energy input will lead to an increased number of pairs, but does not increase the source temperature. \citet{Fabian2015} analyzed all sources observed up to that point by \nustar\ and found nearly all sources lay just below the pair-production limit for thermal Comptonization, suggesting that these coronae are pair-dominated plasmas. 

\begin{figure}
    \centering
    \includegraphics[width=0.9\linewidth]{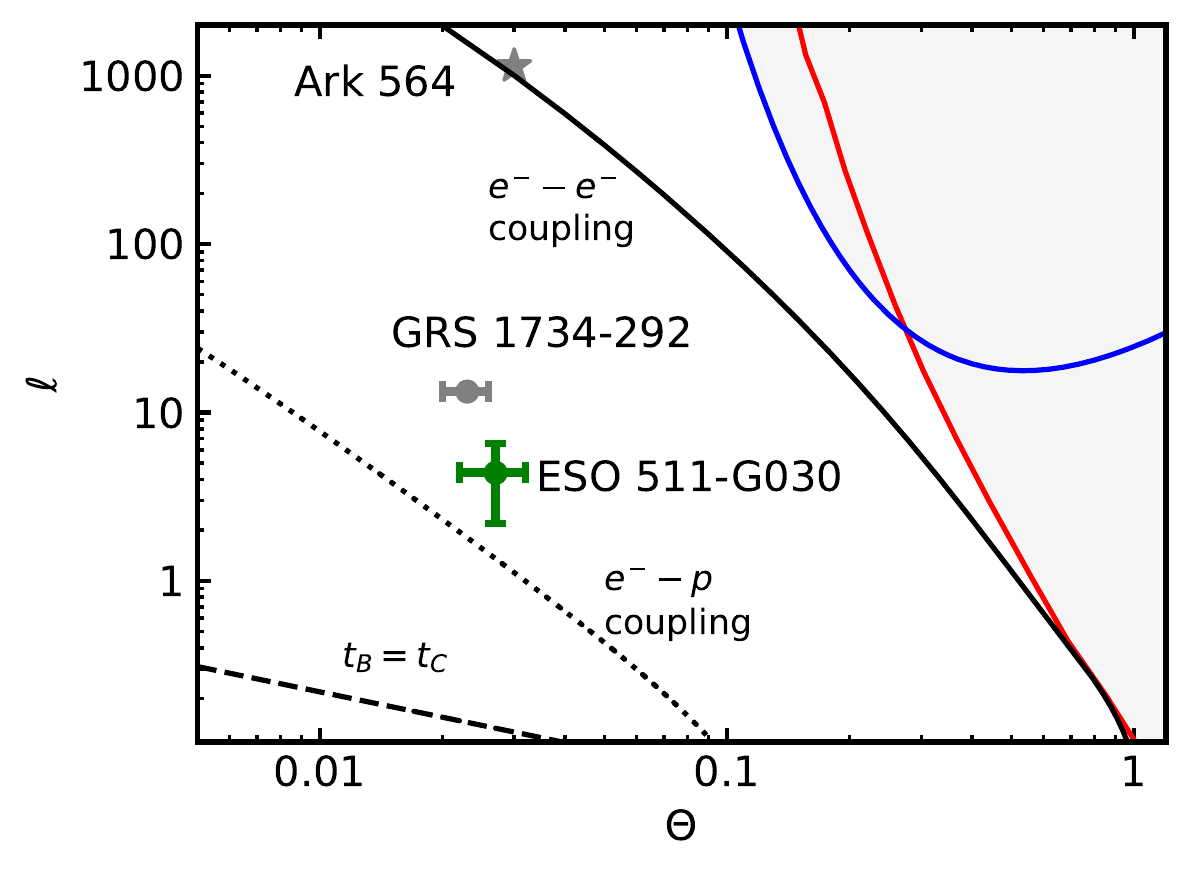} \\
    \caption{{Theoretical compactness–temperature diagram. The green cross indicates the position of \src. The lines are summary of theoretical understanding of the $\ell$-$\Theta$ plane as described in \citet{Fabian2015}; included are the boundaries for electron-electron coupling, electron-proton coupling, the dominance of Compton cooling and pair lines for different assumptions. The positions of GRS~1734-292 and Ark~564 are also indicated in the figure. Located on the similar position, \src\ and GRS~1734-292 would have similar properties.} }
    \label{theta_ell_1}
\end{figure}

\begin{figure}
    \centering
    \includegraphics[width=0.9\linewidth]{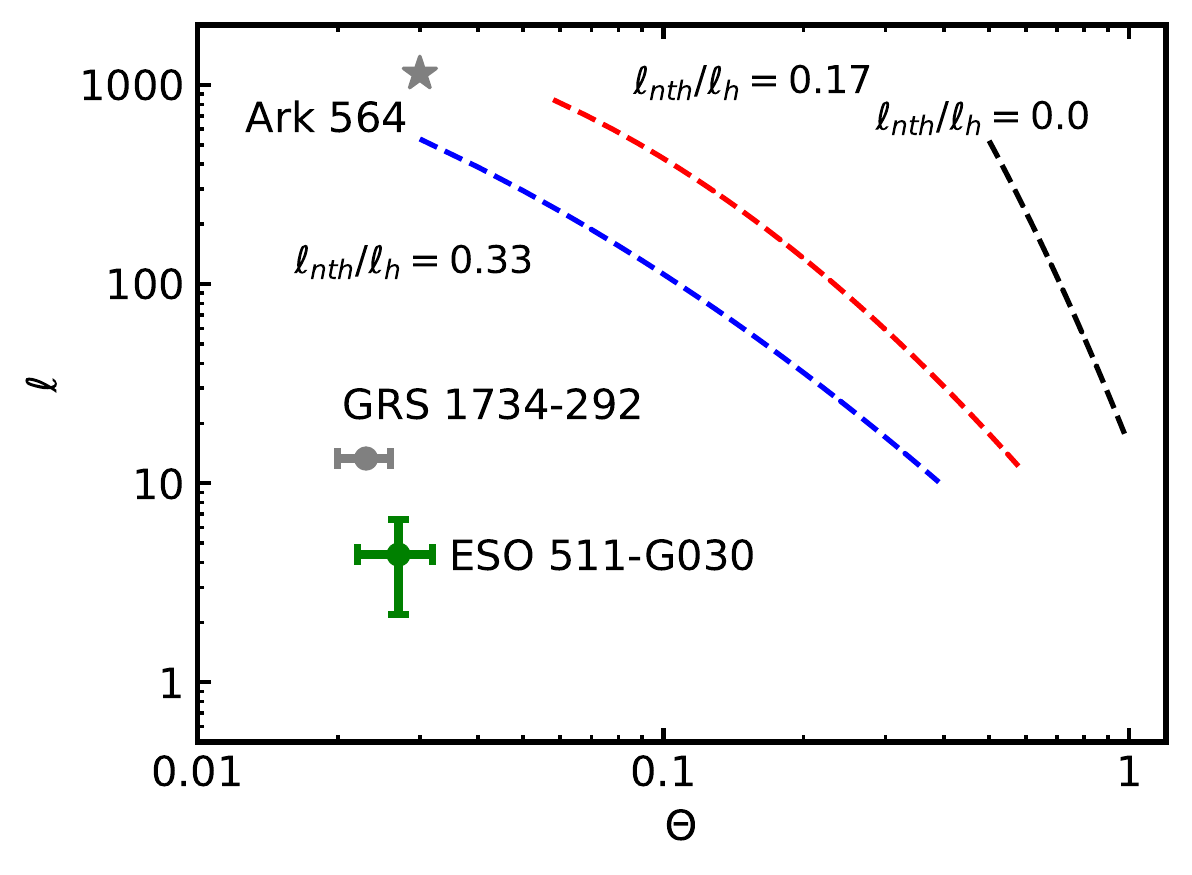} \\
    \caption{{Theoretical compactness–temperature diagram. The green cross indicates the position of \src . The dashed lines are theoretical predictions of a hybrid plasma with $\ell_{\rm h}/\ell_{\rm s} = 1$, where $\ell_{\rm h}$ is the total heating parameter and $\ell_{\rm s}$ is the compactness of the soft photons. The blue, red and black lines represent solutions with a non-thermal fraction ($\ell_{\rm nth}/\ell_{\rm h}$) of 0.33, 0.17 and 0.0 respectively (see \citet{Fabian2017}).} }
    \label{theta_ell_2}
\end{figure}

To compare our results of \src\ with the results of \citet{Fabian2015}, we conduct the same analysis with the result of model~3. With model~3, we get an electron temperature of 13.8~keV ($\Theta=0.027$). Following \citet{Fabian2015}, we measure the Comptonization component from 0.1–200~keV to be $1.95\times10^{43}$~erg~s$^{-1}$. For simplicity, we assume a value of $R = 10R_{\rm g}$, which is a conservative assumption given the measurements. This leads to a compactness of $\ell=4.4$. \src\ lies well below the thermal pair-production limit and above the $e^{-}-p$ coupling line (Fig.~\ref{theta_ell_1}), and thus, assuming a thermal Comptonization model, we find that the corona in this source is a pair-dominated plasma. However, \citet{Fabian2017} re-examined the case of hybrid coronae \citep{Zdziarski1993}, where the plasma contains both thermal and non-thermal particles, as might be expected for a highly magnetized corona powered by the dissipation of magnetic energy. Searching the position in Fig. 6 of \citet{Fabian2017}, we find the corona of \src\ is prone to a hybrid plasma with large fraction of electrons following a non-thermal distribution, {as shown in Fig.~\ref{theta_ell_2}}. Further deep hard X-ray observations are required to distinguish these two scenarios.

Peculiarly, a low coronal temperature is seen in only a handful of AGN, such as 1H0419$-$577 (${kT}_{\rm e}=30^{+22}_{-7}$~keV; \citealt{Jiang2019a}), Ark~564 (${kT}_{\rm e}=15\pm2$~keV; \citealt{Kara2017}), GRS~1734$-$292 (${kT}_{\rm e}=11.9^{+1.2}_{-0.9}$~keV; \citealt{Tortosa2017}), IRAS~13197$-$1627 (${kT}_{\rm e}<42$~keV; \citealt{Walton2018}), 4C~50.55 (${kT}_{\rm e}\approx30$~keV; \citealt{Tazaki2010}), IRAS~04416$+$1215 (${kT}_{\rm e}=3 \sim 20$~keV; \citealt{Tortosa2022}), ESO 362$-$G18 ((${kT}_{\rm e}\approx20$ ~keV; \citealt{Xu2021}) and 3C~273 (${kT}_{\rm e}=12\pm1$~keV; \citealt{Madsen2015b}). Note that except the Seyfert 1 galaxy GRS 1734$-$292 ($\lambda_{\rm Edd}\sim0.03$), the Seyfert 1.5 Galaxy ESO 362$-$G18 ($\lambda_{\rm Edd}\sim0.02$) and the Seyfert 1.8 Galaxy IRAS 13197$-$1627 ($\lambda_{\rm Edd}\sim0.05$$-$$0.1$), all the other sources mentioned above are accreting at a significant fraction of Eddington. In this work, we find another source with a low-temperature corona in a significantly sub-Eddington regime. {For high-Eddington accreting system, the sources would be cooled by relatively cooler seed photons, thus producing the lower temperature cut-off that is observed (e.g., \citealt{Kara2017}). Inversely, the accretion rate of \src\ is only a few per cent of the Eddington limit, so the effectiveness of the cooling mechanism cannot be related to a particularly strong radiation field. However, the high value of the optical depth $\tau$ could be, at least, partly responsible for effectiveness of the cooling mechanism in such sub-Eddington regime (\citealt{Kamraj2022}). Indeed, models predict an anticorrelation between coronal temperature and optical depth (e.g. \citealt{Petrucci2001}). But the reason for the unusually large value of the optical depth is unclear (\citealt{Tortosa2017}).}

{The differences between the possible physical scenarios for the soft excess and low coronal temperatures are significantly much more evident outside the energy coverage and energy resolution of current observatories. With the increasing amount of high-quality spectra from \nustar, it would be possible to start seriously pondering these questions. It would be conclusively distinguished with future missions such as \textit{Athena} (\citealt{Nandra2013}), eXTP (\citealt{Zhang2016}), XRISM (\citealt{Tashiro2018}) and HEX-P (\citealt{Madsen2018}), which are crucial for providing reliable high-resolution X-ray observations. In addition, HEX-P provides a broadband (2–200~keV) response and a much higher sensitivity than any previous mission in the hard energy band. The nature of the corona is likely to be determined by the high-quality observations performed by HEX-P.}


\section{Conclusions}
\label{conclusion}

We investigated the variability and spectra properties from the joint \xmm\ and \nustar\ observing campaign on the Seyfert 1 Galaxy \src. Lightcurve and spectral analysis on its \xmm\ and \nustar\ simultaneous observations show that:

\begin{enumerate}

\item The source remained in a relatively constant flux state throughout the observation period. 
\item The broadband (0.3–78 keV) spectrum shows the presence of a power-law continuum with a soft excess below 2 keV, a relatively narrow iron K$\alpha$ emission ($\sim$6.4 keV), and an obvious cutoff at high energies.
\item The low temperature ($kT_{\rm e}\sim$ 13~keV) of the hot corona are required in all models. {Future X-ray missions (e.g., \textit{ Athena}, eXTP, XRISM and HEX-P) may conclusively distinguish between the possible physical scenarios for the coronae.}
\end{enumerate}


\section*{Acknowledgements}

This work was supported by the National Natural Science Foundation of China (NSFC), Grant No. 11973019, the Natural Science Foundation of Shanghai, Grant No. 22ZR1403400, the Shanghai Municipal Education Commission, Grant No. 2019-01-07-00-07-E00035, and Fudan University, Grant No. JIH1512604. J.J. acknowledges the support from the Leverhulme Trust, the Isaac Newton Trust and St Edmund's College, University of Cambridge.


\bibliography{ESO511}{}
\bibliographystyle{aasjournal}

\end{document}